\documentclass[aps,
  prd,
  onecolumn,
  nofootinbib,
  preprintnumbers,
  superscriptaddress,
  floatfix
]{revtex4-2}

\usepackage[T1]{fontenc}
\usepackage[utf8]{inputenc}
\usepackage{lmodern}

\usepackage{amsmath,amssymb,amsfonts,mathtools}
\usepackage{bm}
\usepackage{hyperref}
\usepackage{graphicx}
\usepackage{caption}
\usepackage{subfig} 
\usepackage{epsfig}
\usepackage{float}
\usepackage{booktabs}
\usepackage{tabularx,array}
\usepackage[dvipsnames]{xcolor}

\begin{document}

\title{Transport coefficients of charged Gauss--Bonnet black holes with arbitrary topology}

\author{Mois\'es Bravo-Gaete}
\affiliation{Departamento de Matem\'atica, F\'isica y Estad\'istica, Facultad de Ciencias B\'asicas, Universidad Cat\'olica del Maule, Casilla 617, Talca, Chile.}
\email{mbravo@ucm.cl, moisesbravog@gmail.com}

\author{Luis Guajardo}
\affiliation{Instituto de Matem\'aticas (INSTMAT), Universidad de Talca, Casilla 747, Talca 3460000, Chile.}
\email{luis.guajardo.r@gmail.com}

\author{Daniel F. Higuita-Borja}
\affiliation{Facultad de Sistemas Biol\'ogicos e Innovaci\'on Tecnol\'ogica, Universidad Aut\'onoma Benito Ju\'arez de Oaxaca, Av. Universidad s/n, C.P. 68120 Oaxaca de Ju\'arez, Oaxaca, M\'exico.}
\affiliation{Instituto de F\'isica, Universidad de Antioquia, Calle 70 No. 52-21, Medell\'in, Colombia.}
\email{dfhiguit@gmail.com}

\author{Julio A. M\'endez-Zavaleta}
\affiliation{Facultad de F\'isica, Universidad Veracruzana, Paseo No. 112, Desarrollo Habitacional Nuevo Xalapa, C.P. 91097, Xalapa-Enr\'iquez, M\'exico.}
\email{julmendez@uv.mx}

\date{\today}

\begin{abstract}
In this study, we present a novel family of exact black hole solutions constructed in the context of five-dimensional Gauss-Bonnet gravity. These solutions add a non-linear charge to the Ba\~nados-Teitelboim-Zanelli-like configurations known to exist with arbitrary Thurston horizon geometry. We establish constraints on the parameter space defining physically viable black holes, aligning with the standard energy conditions. An explicit proof of the first law of thermodynamics within our scenario is provided. We also employ holographic techniques to characterize the DC conductivities for the distinct horizon geometries, identifying a critical temperature indicative of phase transitions and exploring pertinent limits.
\end{abstract}

\maketitle
\flushbottom

\section{Introduction \label{sec:Intro}}

Hawking's theorem  \cite{Hawking:1971vc} and topological censorship \cite{Friedman:1993ty},  unassailably determine the horizon topology of stationary four-dimensional black holes (BH) to be spherical when the dominant energy condition (DEC) is satisfied. The task of bypassing these results has materialized itself in new BH configurations with non-trivial topology, either (i) in the framework of higher-dimensional gravity or (ii) considering deviations from the DEC.

 Concerning point (i), in five dimensions, one finds the emblematic example of  \textit{black ring} solutions.
 In \cite{Emparan:2001wn}, Emparan and Reall showed that the vacuum Einstein equations admit a stationary and asymptotically flat spacetime with an event horizon of topology $\mathbb{S}^{1} \times \mathbb{S}^{2}$.
 Black rings exemplify the breakdown of uniqueness in five-dimensional gravity. 
 They coexist with the well-known Myers-Perry BH \cite{Myers:1986un}, which allocates a $\mathbb{S}^{3}$ horizon topology. 
On the other hand, the violation of the DEC (ii) can be performed via the inclusion of a cosmological constant $\Lambda$ in the Einstein-Hilbert (EH) action, allowing one to obtain the denominated topologically Anti-de Sitter  (AdS)  BHs (see, for example, Refs.~\cite{Lemos:1994xp,Lemos:1995cm,Birmingham:1998nr,Aminneborg:1996iz}). Therein, the event horizon's topology is parametrized by the constant spatial curvature, which can manifest as a four-dimensional sphere (positive curvature), a torus (zero curvature), or a hyperbolic space (negative curvature).

In particular, for a five-dimensional spacetime, the event horizon of a stationary BH is a ruled hypersurface foliated by a compact orientable three-dimensional Riemannian manifold \cite{Cadeau:2000tj}.  
According to Thurston's conjecture \cite{thurston1997three,scott1983geometries}, proved by Perelman in \cite{perelman2002entropy,perelman2003ricci}, any compact orientable three-dimensional Riemannian manifold can be decomposed into products of the geometries showcased in Tab.~\ref{tab:Thurston}:
\begin{table}[H] 
\begin{tabularx}{\textwidth}{ @{\extracolsep{\fill}}
  l l l
  >{$\displaystyle} c <{\vphantom{\sum_{1}{N}}$} >{\refstepcounter{equation}(\theequation)} r
  @{} }   \toprule
\textit{Topology} & \textit{Line element} & \multicolumn{1}{l}{\textit{Bianchi Type}} & \multicolumn{1}{l}{\textit{Value of K}} & \multicolumn{1}{l}{} \\ \midrule	
$\mathbb{E}^{3}$ & $d\tilde{s}^2=dx_{1}^{2}+dx_{2}^{2}+dx_{3}^{2}$ & $\text{I}$ & 0 & \label{euclidean}   \\
$\mathbb{S}^{3}$  &  $d\tilde{s}^2=\sin^2(x_3)\left(\sin^2(x_2)dx_{1}^{2}+dx_{2}^{2}\right)+dx_{3}^{2}$ & $\text{IX}$ & 6 & \label{sphere} \\
$\mathbb{H}^{3}$  &  $d\tilde{s}^2=\sinh^2(x_3)\left(\sin^2(x_2)dx_{1}^{2}+dx_{2}^{2}\right)+dx_{3}^{2}$ & $\text{V}$ & -6 & \label{hyperbolic} \\
$\mathbb{S}^{1} \times \mathbb{S}^{2}$ &  $d\tilde{s}^2=dx_{1}^{2}+dx_{2}^{2}+\sin^{2}(x_{2})\,dx_{3}^{2}$ & $\text{--}$ & 2 & \label{S1timesS2} \\
$\mathbb{S}^{1} \times \mathbb{H}^{2}$ & $d\tilde{s}^2=dx_{1}^{2}+dx_{2}^{2}+\sinh^{2}(x_{2})\,dx_{3}^{2}$ & $\text{III}$ & -2 & \label{S1timesH2} \\
Nil   &   $d\tilde{s}^2=dx_1^2+dx_2^2+(dx_3-x_1dx_2)^2$ & $\text{II}$  & -\tfrac{1}{2} & \label{Nilg}   \\
Solv  &   $d\tilde{s}^2=x_3^{2}dx_1^2+\frac{1}{x_3^{2}}dx_2^2+\frac{1}{x_3^{2}} dx_3^2$ & $\text{VII}_0$ & -2 & \label{Solvg}  \\ 
$\mbox{SL}_2(\mathbb{R})$   &   $d\tilde{s}^2=
\frac{1}{x_1^2}(dx_1^2+dx_2^2)+\Big(dx_3+\frac{dx_2}{x_1}\Big)^2$ & $\text{VIII}$ & -\tfrac{5}{2} & \label{Sl2}  \\
 \bottomrule
 \hline
\end{tabularx}
 \caption{\label{tab:Thurston}  The eight Thurston spaces that fulfill the geometrization conjecture,  as well as their metric tensors and their Bianchi type. Recall that $\mathbb{S}^2\times \mathbb{S}^1$ cannot be realized as a left-invariant metric. For later purposes, we also include their scalar curvature $K$.}
\end{table}
\begin{table}[H]\centering
\begin{tabular}{@{} *4l @{}}    \toprule
\textit{Horizon topology} & \textit{Supporting matter fields}& \textit{Rotation}  &  \textit{Reference}
\\\midrule
Nil and Solv  & Maxwell field  &   No  & \cite{Arias:2017yqj}   \\
 Nil and Solv  & Dyonic dilaton-Maxwell fields &  No &  \cite{Bravo-Gaete:2017nkp} \\
 Solv & Magnetic gauged $U(1)$ supergravity  & No  &  \cite{Faedo:2019rgo}\\
Solv & None &  Slowly rotating & \cite{Figueroa:2021apr} \\
 Nil and $\mathbb{S}^{1} \times \mathbb{H}^{2}$ & $z=4$ Ho\^rava-Lifshitz term & No  & \cite{Naderi:2021skd}  \\
  Nil, Solv and $\mbox{SL}_2(\mathbb{R})$  & Dyonic dilaton-$B$-field  &   No  & \cite{Naderi:2019jhn}   \\
   Nil and $\mbox{SL}_2(\mathbb{R})$  & Chern-Simons term  &   Yes  & \cite{Faedo:2022hle}   \\
   Nil, Solv and $\mbox{SL}_2(\mathbb{R})$  & Scalar field non-minimally coupled  &   No  & \cite{Guajardo:2024hrl}   \\
 \bottomrule
 \hline
\end{tabular}
 \caption{\label{tab:LitBHs} Some five-dimensional BH solutions in the literature with a non-trivial horizon topology.  }
\end{table}

 In Tab.~\ref{tab:Thurston}, $\mathbb{E}^{3}, \mathbb{S}^3$ and $\mathbb{H}^{3}$ are the maximally symmetric spaces, corresponding to zero, positive, and negative curvatures, respectively.
 In addition to the Cartesian products $\mathbb{S}^{1} \times \mathbb{S}^{2}$ and $\mathbb{S}^{1} \times \mathbb{H}^{2}$, the last three are non-trivial manifolds known in the literature as Nil geometry, Solv geometry, and the geometry of the universal cover of $\mbox{SL}_{2}(\mathbb{R})$.
Altogether they provide the exhaustive classification of three-dimensional manifolds and, consequently, the possible topologies for compact event horizons.
Following this line, a few years ago Cadeau and Woolgar \cite{Cadeau:2000tj} were the first to find new families of BH solutions whose horizons are described by the Nil (\ref{Nilg}) and  Solv (\ref{Solvg}) geometries. In analogy to this pioneering work, novel five-dimensional configurations were reported.
 For example, in vacuum Einstein gravity, a solution exhibiting hyperscaling violation in Nil geometry (\ref{Nilg}) was described in \cite{Hassaine:2015ifa}. Charged BHs for Nil and/or Solv geometries, governed by the Einstein-Hilbert action  (EH) (with or without a cosmological constant  $\Lambda$) are also documented in the literature. Some of these solutions require supporting matter fields. In Tab.~\ref{tab:LitBHs}, we provide a brief overview of these solutions.

Similar to the matter-sourced EH-$\Lambda$, we may consider models containing higher-curvature corrections for the same objective. Among such models, those preserving covariance and second-order equations of motion are referred to as Lanczos-Lovelock gravity  \cite{Lanczos:1938sf,Lovelock:1971yv},  where the  EH-$\Lambda$ action is contained as a particular case. A key characteristic is that the whole theory remains free of ghostly pathologies \cite{Zwiebach:1985uq,Zumino:1985dp}. 
In five dimensions, the simplest non-trivial extension is captured via the gravity action
{\begin{align}\label{eq:GBaction}
 S_{\mbox{\tiny{GB}}}=\frac{1}{2\kappa} \int{  d^5x \sqrt{-g}  \left( { R-2\Lambda}+{\alpha} \,\mathcal{G} \right)},
\end{align}}
where
  \begin{align} \label{eq:GBterm}
 \mathcal{G}=R^2-4R^{\mu\nu}R_{\mu\nu}+R^{\alpha\beta\mu\nu}R_{\alpha\beta\mu\nu},
 \end{align}
is dubbed as the Gauss-Bonnet term (GB) in the literature. Here, $\kappa$ is a constant related to the Newton constant,  while  $\alpha$ corresponds to the GB coupling constant. The equations of motion read 
 \begin{align} \label{eq:eqmotion}
 \mathcal{G}_{\mu \nu}:=&\, G_{\mu \nu} +\Lambda g_{\mu \nu}
+ \alpha \Big( -4 R_{\mu}^{\phantom{\sigma} \sigma} R_{\nu \sigma} + 
 2 g_{\mu \nu} R_{\sigma \rho}R^{\sigma \rho} + 
 2 R_{\mu \nu}R  &\\&- 4 R^{\sigma \rho} R_{\mu \sigma \nu \rho} + 
 2 R_{\mu}^{\phantom{\mu} \sigma \rho \tau} R_{\nu \sigma \rho \tau } - \frac{1}{2} g_{\mu \nu} \mathcal{G}\Big) =0. \nonumber
 \end{align}
It is noteworthy to emphasize the rich variety of results stemming from the theory (\ref{eq:GBaction})-(\ref{eq:GBterm}). Notable instances include analytic  spherically symmetric solutions \cite{Boulware:1985wk}, topological  BHs \cite{Cai:2001dz,Cai:1998vy}  and charged and hairy configurations  \cite{Cvetic:2001bk,Bravo-Gaete:2021hza,BravoGaete:2013acu}. One particularly intriguing scenario within the Einstein-GB model  \eqref{eq:GBaction} occurs when the relation
\begin{equation}
\label{rel_lambda}
4\Lambda \alpha = -3,
\end{equation}
is met.   At this particular choice of coupling, we say that the theory sits at the \emph{Chern–Simons point} (hereafter, the CS point). As a consequence of \eqref{eq:GBaction}, the action transforms into a topological gauge theory of gravity, based on the CS five-form and the $SO(1,5)$ gauge group \cite{Chamseddine:1989nu}. 
In this regard, it is worth recalling that generic Lovelock terms present bifurcated vacuum branches, not all of which are physically acceptable: linear fluctuations around some of them typically reveal the presence of Boulware–Deser ghosts. At the CS point, however, the two branches associated with the GB term degenerate into a unique AdS vacuum, thereby eliminating the ghostly branch. This special feature leads to a consistent classical theory in which the linear propagator becomes degenerate, thus avoiding the emergence of unphysical degrees of freedom and preserving unitarity \cite{Crisostomo:2000bb}. Under these conditions, the model accommodates BH solutions, as reported in \cite{Dotti:2007az}, notable for the striking property that \emph{any} base manifold is compatible with the field equations. The  BH configurations considered in this work reside precisely at the CS point.

It is natural to ask for charged extensions of the aforementioned BHs. In this context, non-linear sources have been intensively explored in recent years. The Born-Infeld theory and the Euler-Heisenberg effective theory (in the quantum electrodynamic case) paved the way to explore
new models beyond Maxwell. Detouring from the classical Maxwell action has unveiled new avenues of exploration in BH physics, inflation, and dark energy problems. Also,  as descriptions
of certain condensed matters and optical media. Notably, recent studies have yielded intriguing findings, such as the behavior of the Joule-Thompson expansion in Quasi-topological Electromagnetism  \cite{Barrientos:2022uit}, and BHs in ModMax theory \cite{Barrientos:2022bzm,Barrientos:2025rde}. We refer the reader to \cite{Sorokin:2021tge} for a recent review on the topic. 

In this work, under the foundational hypothesis of a completely arbitrary base geometry, we explore Pleb\'anski's $H(P)-$ non-linear electrodynamics as a source for Eqs. (\ref{eq:GBaction})-(\ref{eq:GBterm}),  modeled by the action
\begin{align}\label{eq:NLE}
S_{\mbox{\tiny{NLE}}} = \int{d}^5x\sqrt{-g}\, {\cal{L}}_{\mbox{\tiny{NLE}}}=\int{d}^5x\sqrt{-g}\Biggl(-
\frac{1}{2}F_{\mu\nu}P^{\mu\nu} + {H}(P)\Biggr).
\end{align}

The basic ingredient to cast the Lagrangian density of a non-linear Maxwell source $\mathcal{L}=\mathcal{L}(F)$ is the scalar $F=(F_{\mu\nu}F^{\mu\nu})/4$. This quantity is constructed out of the Maxwell tensor $F_{\mu\nu}=\partial_{\mu}A_{\nu} - \partial_{\nu}A_{\mu}$. However, this prescription accepts a first-order rewriting in terms of the Pleb\'anski tensor $P_{\mu\nu}$, which is formally defined through
\begin{equation}
P_{\mu\nu} \equiv 2\left(\dfrac{\partial \mathcal{L}}{\partial F^{\mu\nu}}\right) = \mathcal{L}_F F_{\mu\nu},
\end{equation}
where $\mathcal{L}_F$ denotes the partial derivatives of the Lagrangian density with respect to $F$. Via a Legendre transformation, the Lagrangian can be successfully rendered in the form  \eqref{eq:NLE} using a structural function ${H(P)}$, formally a Hamiltonian of the theory, which now depends on the invariant  $P=(P_{\mu\nu}P^{\mu\nu})/4$. A succinct treatise on the topic can be found in \cite{Plebanski:1970zz}. Among other things, this formalism has proven to be a powerful tool to find completely regular \cite{Ayon-Beato:1998hmi,Ayon-Beato:2000mjt,Ayon-Beato:1999kuh}, rotating \cite{DiazGarcia:2022jpc,Garcia-Diaz:2021bao,Ayon-Beato:2022dwg,Bravo-Gaete:2025xya}, AdS \cite{Alvarez:2022upr,Lin:2024ubg}, charged and hairy \cite{Bravo-Gaete:2022mnr,Hyun:2019gfz,Churilova:2019sah}  BHs.

As we will see below, the electrically charged solution will require, as an extra condition, that the base manifold possess a constant scalar curvature, denoted by $K$, which in turn is the case for the Thurston geometries (see Tab. \ref{tab:Thurston}). A relaxed fall-off to the vacuum solution \cite{Dotti:2007az} ensures non-vanishing and finite charges. In the realm of holography, it is well established that Reissner-Nordstr\"{o}m BHs featuring translation invariance lead to infinite conductivity, as charge carriers cannot dissipate their momentum due to this invariance \cite{Hartnoll:2009sz}. Therefore, one method to achieve finite conductivities and delve into new physics is by explicitly breaking this invariance.
Supplementing the theory with additional matter fields has proven to be a fruitful route in this regard (see, e.g., the study of lattices \cite{Horowitz:2012ky, Horowitz:2012gs, Donos:2012js, Donos:2013eha} and axionic models \cite{Andrade:2013gsa, Cisterna:2018hzf, Cisterna:2019uek, Hao:2022zkr,BravoGaete:2019rci}).
A different path results from contemplating anisotropic spacetimes, such as Nil and Solv geometries \cite{Arias:2017yqj}, in which the dissipation mechanism arises not from the interaction with matter fields but rather as a geometric consequence. Here, we analyze the DC-conductivity of charged Gauss-Bonnet BHs whose event horizon is governed by non-maximally symmetric Thurston geometries, further reinforcing the idea that anisotropic spaces offer a natural mechanism for momentum relaxation. For the non-standard cases \eqref{S1timesS2}--\eqref{Sl2}, we construct a unique non-linear electrodynamics source that supports electrically charged BHs within these geometries. The uniqueness of the electric source is established by employing the Plebánski $H(P)$-formalism of non-linear electrodynamics.

 In addition to the geometrical novelty of the solutions presented, our analysis uncovers an intriguing transport behavior. In particular, we identify the emergence of critical temperatures and negative branches in the DC conductivity profiles. Although such anomalous transport phenomena might seem unexpected at first glance, they resemble effects that have been well-documented in strongly driven condensed matter systems, like semiconductor superlattices and two-dimensional electron gases under microwave irradiation \cite{Bykov2008,PhysRevB.68.115324,RYZHII200413, Ryzhii2004}. This surprising connection not only highlights the physical richness of the solutions we present but also opens a way to model these anomalous critical phenomena from a holographic perspective.
\\\,\\
\emph{Organization of the Manuscript}. The content of this work is presented in the following order: In Sec. \ref{sec:BlackHole}, we introduce a novel electrically charged BH for the Einstein-GB-$H(P)$ model, featuring a horizon geometry ruled by any of the Thurston geometries, thereby generalizing the solutions of \cite{Dotti:2007az}. In Sec. \ref{sec:EnergyCon}, we analyze the energy conditions on the non-linear source, demonstrating that the standard conditions can be met by imposing a lower bound on the configurations's mass. The thermodynamics of the global charges are constructed in Sec. \ref{sec:Thermo}, and taking advantage of these results, we delve into the analysis of holographic DC conductivity in Sec. \ref{sec:Transport}. In Sec. \ref{sec:interpretation}, we analyze the physical interpretation of the identified anomalous transport regimes, emphasizing their qualitative resemblance to known non-equilibrium phenomena. Finally, we conclude with relevant discussions and ponder potential avenues for further lines of investigation in Sec. \ref{sec:Remarks}.

\section{Five-Dimensional {charged} black holes \label{sec:BlackHole}}
 
Classical extensions to exact gravitational solutions stem mainly from three sources:  relaxation of symmetry considerations, alternative gravitational dynamics, or the addition of matter fields. In either case, one may end up with a solution manifesting additional global charges, thus turning its physics more enticing. 
 For instance, following the last option, it is customary to add a Maxwell field, which naturally leads to the incorporation of a $U(1)$ charge. As we stated in Sec.~\ref{sec:Intro}, we follow this approach with a slightly more ambitious scope. 
 Supplementing the Einstein-GB action (\ref{eq:GBaction})-(\ref{eq:GBterm}), we will incorporate the non-linear electrodynamics described by the $H(P)-$formalism (\ref{eq:NLE}). Consequently, the action under consideration is as follows:
\begin{equation}\label{eq:action}
	S= S_{\mbox{\tiny{GB}}}+S_{\mbox{\tiny{NLE}}}.
\end{equation}
We confine the focus of this work to the search for purely electrical and static configurations. 
Under such circumstances, the variation of the action (\ref{eq:action}) leads to field equations 
\begin{subequations}\label{eq:EOM}
\begin{align}
\nabla_{\mu}P^{\mu\nu}& =0, \label{eq:Maxwell}\\
F_{\mu\nu}&=\left(\frac{\partial {H}}{\partial P}\right) P_{\mu \nu}= {H}_P P_{\mu\nu}\,
\label{eq:constitutive},\\
 \mathcal{E}_{\mu \nu}&:= \mathcal{G}_{\mu \nu}-\kappa {T}_{\mu\nu}=\mathcal{G}_{\mu \nu}-\kappa \left({{H}_P} P_{\mu\alpha}P_{\nu}^{\, \alpha}
-g_{\mu\nu}(2P {H}_P-{H})\right)=0,\label{eq:TmunuNLE1}
\end{align}
\end{subequations}
with $\mathcal{G}_{\mu \nu}$ previously defined in eq.~(\ref{eq:eqmotion}).

As demonstrated in \cite{Dotti:2007az}, BHs for the vacuum configuration can be obtained through the following ansatz:
\begin{align} \label{eq:metric}
 ds^{2}=-f(r) dt^{2}+\frac{dr^2}{f(r)} + r^{2} h_{i j} dx^{i} dx^{j},
\end{align}
where $-\infty<t<+\infty$, $r \geq 0$ and the $h_{i j}$'s depend on the spatial coordinates, with $i,j,k,\dots \in \{x_1,x_2,x_3\}$. 

Despite the arbitrary horizon topology $h_{i j}$, the combination of Einstein equations $\mathcal{E}_2^2-\mathcal{E}_3^3$ leads to the simple relation \begin{align}\left(-\frac{1}{2} +\alpha f''\right) \Xi =0,
\label{eq:int_cond}\end{align}
where $\Xi$ is an expression that depends on the functions $h_{i j}$, their first partial derivatives ($\partial_{k} h_{i j}$) and the second derivatives $\left(\partial^2_{kl} h_{i j}\right)$. 
In the special case of maximally symmetric spaces \eqref{euclidean}-\eqref{hyperbolic}, $\Xi\equiv0$ trivially.
 As a result, eq. \eqref{eq:int_cond} offers no information for integrating either the metric or the matter fields. Specifically, the function $f$ remains entirely arbitrary, leading to a configuration of no physical relevance, as will be discussed later on. 
 This arbitrariness is the geometric manifestation of the CS point vacua degeneracy: around constant curvature spaces, it is well-established that the linearized Einstein-GB equations trivialize, meaning that no sensible graviton propagator exists, see for instance  \cite{Fan:2016zfs}. We emphasize that this issue at the linear level is a characteristic underlying the whole Lanczos-Lovelock family, exclusively when probing metric perturbations on maximally symmetric backgrounds \cite{Devecioglu:2024qiu}. 
Whether through this trivialization or the more intriguing scenario in which the metric function satisfies the constraint
 $$-\dfrac{1}{2} + \alpha f'' = 0,$$ 
the mixed form of the  Einstein equations $\mathcal{E}_{\mu}^{\ \nu}$ becomes diagonal, taking the form $\left(\mathcal{E}_{\mu}^{\ \nu}\right) = \text{diag}(N,N,S,S,S)$, regardless of the base manifold. Here, $N$ and $S$ represent lengthy expressions that are not displayed as they do not offer any particularly significant information.

It is well-known that vacuum solutions will restrict the gravitational potential to have a  Ba\~{n}ados-Teitelboim-Zanelli (BTZ)-like form. In the presence of matter, the metric function $f(r)$ can allocate a linear term, relaxing its fall-off at infinity. This behavior is indeed possible in our setup by the appropriate choice of the structural function $H(P)$ to be
\begin{equation} \label{eq:HP_sol}
{H}(P)= \dfrac{\alpha \lambda }{\kappa}\sqrt{-2P} - \dfrac{6 \alpha \lambda^{\frac{10}{3}} }{\kappa} \sqrt[3]{-2 P},
\end{equation}
where $\lambda$ is a coupling constant of the nonlinear electrodynamics model. This particular electrodynamics enables constructing an electrically charged  BH stemming from the ansatz (\ref{eq:metric}).  For the sake of completeness, the above structural function can be expressed via the $\mathcal{L}(F)$ formalism and presented in  Appendix \ref{sec:appA}. The arbitrariness of the geometry is preserved modulo the demand of a base manifold with constant scalar curvature, denoted by $K$ as before.  We recapitulate the values of the constant scalar curvature $K$ in Tab. \ref{tab:Thurston} for each of the Thurston geometries.
\noindent
Altogether, the solution for any non-maximally symmetric base manifold is uniquely determined to be
\begin{eqnarray}
f(r) &=& \dfrac{r^2}{4\alpha} + \lambda^2 \sqrt{K+6 \mu}\, r -\mu,\label{eq:solnf}
\\ F_{rt} &=& \frac{4 \lambda^3\alpha r}{\kappa \sqrt{K+6 \mu} }-\frac{\alpha \lambda}{\kappa},\label{eq:solnFrt}
\end{eqnarray}
with $\Lambda$ and $\alpha$ related as in (\ref{rel_lambda}), and $\mu$ is an arbitrary integration constant. For the maximally symmetric cases $\mathbb{E}^{3}, \mathbb{S}^3$ and $\mathbb{H}^{3}$, the metric function $f(r)$ indeed remains arbitrary and is therefore a pure gauge. However, choosing it differently from eq. \eqref{eq:solnf} necessitates a modification of the electrodynamics described by eq. \eqref{eq:HP_sol}, which in turn alters the global charges and the underlying theory. As a result, the construction becomes physically irrelevant. It is worth noting that the field equations (\ref{eq:EOM}) uniquely fix the structural function (\ref{eq:HP_sol}) when we require the existence of an exact electrically charged solution (\ref{eq:solnFrt}) with a linear contribution to the metric function (\ref{eq:solnf}) and an arbitrary constant-curvature base manifold given by (\ref{S1timesS2})-(\ref{Sl2}). Alternative nonlinear electrodynamics models would not support such configurations.

Concerning the configuration determined by the function $f$ as presented in (\ref{eq:solnf}), numerous observations can be made.
First of all, the limit $\lambda=0$ leads to the uncharged configuration found in \cite{Dotti:2007az}. 
Furthermore, it's worth noting that the scalar curvature $R$ for such a class of spacetimes is expressed as
\begin{eqnarray}\label{eq:R}
	R=-\frac{6f'}{r}-f''+\frac{K}{r^2}-\frac{6f}{r^2},
\end{eqnarray}
where ($')$ denotes the derivative with respect to the coordinate $r$ henceforth.  Thus, upon evaluation of the solution for $f$, we obtain
\begin{eqnarray}\label{eq:Ricciscalar}
	R=\frac{K+6\mu}{r^2}-\frac{12\lambda^2\sqrt{K+6\mu}}{r}-\frac{5}{\alpha},
\end{eqnarray}
showing a curvature singularity $r_s$ located at $r_s=r=0$. 

\begin{figure}[H]
  \centering
    \includegraphics[scale=0.123]{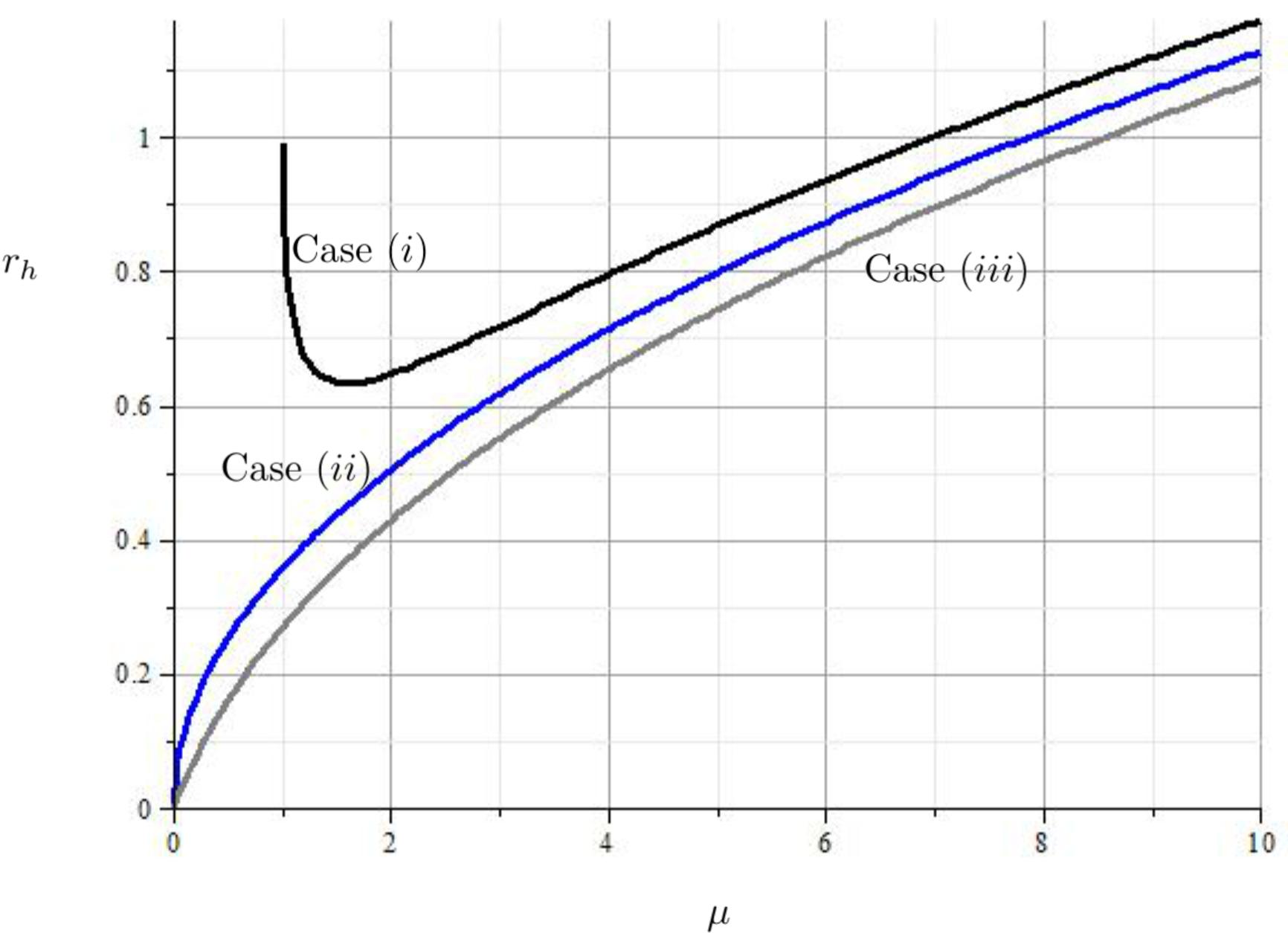}
	\includegraphics[scale=0.11]{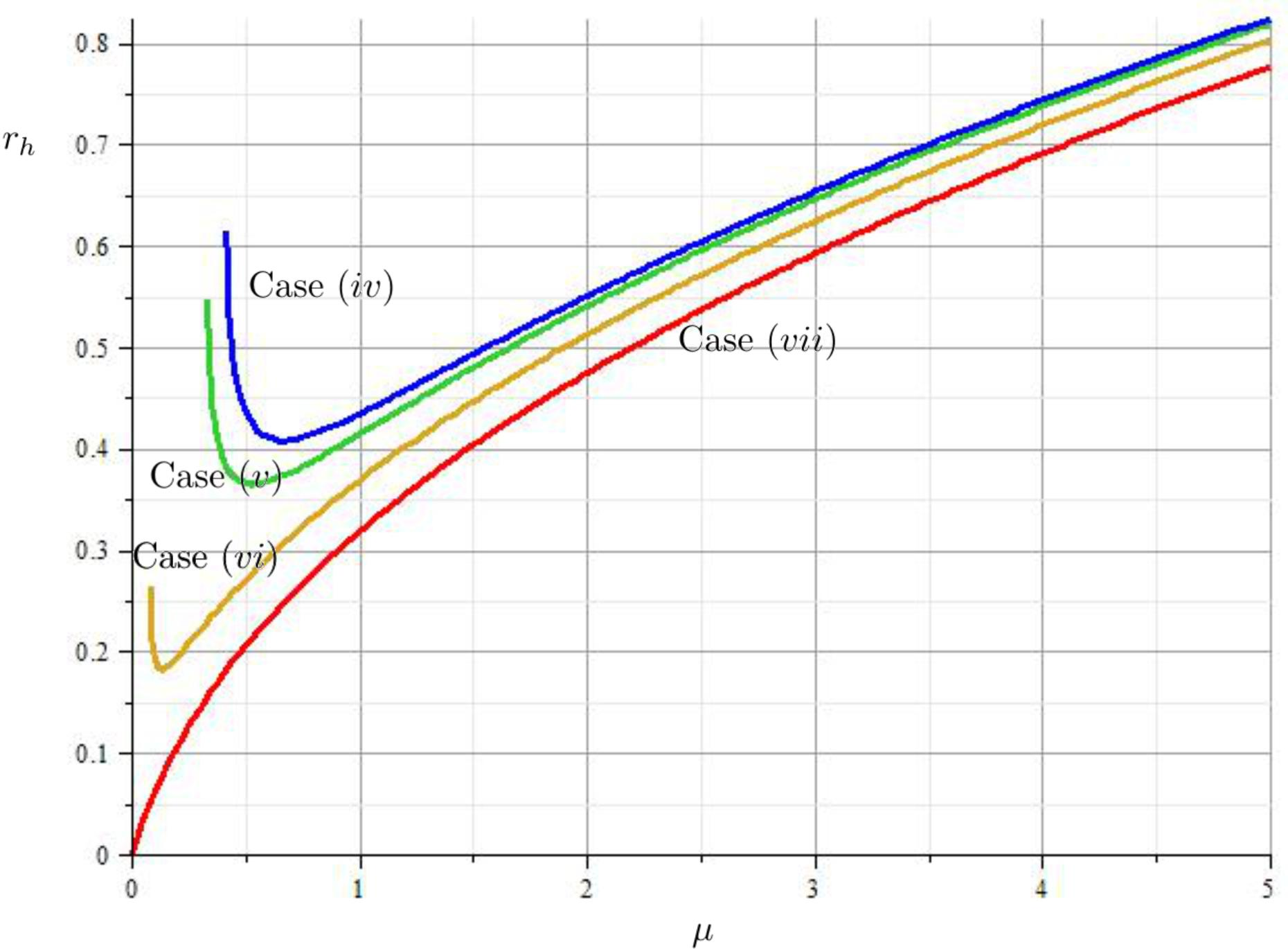}
      \caption{\label{fig:rhmu} Left Panel: Representation of the location of the event horizon $r_h$ depending on the integration constant $\mu$. Here, from Table \ref{tab:Thurston}, the case $(i)$ ($K=-6$) is represented by a black curve, while cases $(ii)$ ($K=0$) and $(iii)$ ($K=6$) correspond to the blue curve and gray curve respectively. Right Panel: Representation of $r_h$ with respect $\mu$, where the case $(iv)$ ($K=-5/2$) is represented by a blue curve, while that cases $(v)$ ($K=-2$) and $(vi)$ ($K=-1/2$) correspond to the green curve and orange curve respectively.  Finally, case $(vii)$ ($K=2$) is visualized via the red curve. Here, we have considered $\lambda=4 \alpha=1$ for our computations.}
\end{figure}

The natural question is whether the function $f$ effectively defines a  BH configuration.  To answer this query, we require that the radius $r_s$ remains concealed within the location(s) of the event horizon(s), denoted as $r_h > 0$, where $f(r_h) = 0$. A thorough examination of the gravitational potential derived from (\ref{eq:solnf}) is providential for such an investigation. Here,  we note that in the asymptotic regime as $r \rightarrow +\infty$, the function  $f(r)  \simeq r^{2}/(4 \alpha)$. Given that $\alpha>0$, $f$ exhibits unbounded growth. Conversely,  when $r \rightarrow +0$, we have that $f(r) \rightarrow -\mu$. A concise examination of the first and second derivatives
\begin{eqnarray}\label{eq:fprimes}
	f'(r)=\frac{r}{2\alpha}+\lambda^2 \sqrt{K+6 \mu}>0,\qquad f''(r)=\frac{1}{2\alpha}>0,
\end{eqnarray}
shows that (\ref{eq:solnf}) is an increasing function that displays a unique event horizon with location  $r_h$, if we suppose that the integration constant $\mu$ is positive, together with $K+6\mu>0$. The explicit value of $r_h$, as a function of the parameter 
$\mu$ keeping $\alpha,\lambda$ and $K$ fixed can be obtained from the positive branch of (\ref{eq:solnf}) 
\begin{equation}\label{eq:rh}
r_h=r_h(\mu)=-2\lambda^2 \alpha  \sqrt{K+6\mu} \pm 2\sqrt{\lambda^4\alpha^2(K+6\mu)+\mu\alpha}.
\end{equation} 
Observe that a positive value $r_h>0$ is assured by the condition $\mu \alpha>0$. This condition remains consistent if, as previously assumed, $\mu$ is a positive integration constant and $\alpha > 0$. It is interesting to note that from (\ref{eq:rh}) and having $r_h>0$, we can obtain a minimum local expression for the location of the event horizon (denoted as $r^{*}_h$), granted by
$$r^{*}_h=\sqrt{-\frac{2\alpha K}{3(6 \lambda^4\alpha+1)}}>0,$$
when the integration constant $\mu$ is fixed as
$$\mu^{*}=-\frac{K (12\lambda^4\alpha+1)}{6(6\lambda^4\alpha+1)}>0.$$
For our analysis, these values are only feasible if $K<0$. From Table \ref{tab:Thurston}, the relationship between $\mu$ and $r_h$ is graphically depicted in Figure \ref{fig:rhmu}.

\section{Energy conditions \label{sec:EnergyCon} }

A fundamental motivation of this work is to walk away from the reach of classical theorems akin to Hawking's topology theorem to pursue non-standard  BH configurations. A consequence is that some physical properties guaranteed by these theorems are no longer certain. Energy positivity, for instance, is a key concern, in particular for later applications.

In our specific situation, two factors might play a role in restraining the direct applicability of known results regarding energy conditions. One of them is the modification of gravitational dynamics, and the other is due to the higher-dimensional nature of the solution. Anyhow, a characterization of the energy conditions problem adapted to our case has been developed over the past years. We opt for the construction detailed in \cite{Maeda:2018hqu}, which, in turn, is grounded on previous approaches \cite{Santos:1994cs, Santos:1995kt, Hall:1995uj}.  We commit this section to determining, under this light, the conditions on the solution (\ref{eq:metric})-(\ref{eq:HP_sol}) needed to render it physically admissible.  

The strategy in \cite{Maeda:2018hqu} extends the well-known Hawking's classification of the energy-momentum tensor originally specialized to four dimensions. The algebraic structure of any $D\geq3$ energy-momentum tensor, with $D$ the dimension of the spacetime, falls into one of the so-called \emph{Segre classes}, which are determined by the multiplicity of its eigenvalues and the causal type of the associated eigenvectors. A preparation step is required to work out an invariant characterization for arbitrary dimension: the eigenvalue problem must be performed on the orthonormal components of the energy-momentum tensor. This is, an orthonormal frame (vielbein) is to be constructed for the background and transform with it the spacetime components to the orthonormal ones.

A vielbein for the BH metric in eq.~\eqref{eq:metric} can be obtained by taking advantage of the special algebraic structure of its line element. The base manifold can be split as the Cartesian product between the Lorentzian space spanned by $(t,r)$ and the $3$-dimensional Euclidean part spanned by the $x^i$ coordinates (with $i=1,2,3$). Accordingly, the complete $D=5$ metric acquires a direct sum form 
$$  g=g^{(2)}\oplus r^2 h \,\,\iff \,\, g_{\mu\nu}dx^{\mu}dx^{\nu}= -f(r)dt^2+\frac{1}{f(r)} dr^2+r^2 h_{ij}dx^{i} dx^{j}.
$$
The frame $e^{(A)}_{\mu}$ is defined, up to Lorentz transformations, through the orthonormality conditions
 \begin{equation}
 	e^{(A)}_{\mu}e^{(B)}_{\nu}\eta_{AB}=g_{\mu\nu}, \qquad e^{(A)}_{\mu}e^{(B)}_{\nu}g^{\mu\nu}=\eta^{AB},
 \end{equation}
where the capital indices  $A,B\in \{0,1,2,3,4\}$ run over the orthonormal basis and $\eta_{AB}$ is the flat Minkowski metric defined therein. It can be shown that $e^{(A)}_{\mu}$ also inherits a direct sum structure, and hence a solution reads 
\begin{align}
e^{(0)}_{\mu}&=\sqrt{f}\delta^{t}_{\mu}, \,\quad e^{(1)}_{\mu}=\frac{1}{\sqrt{f} }\delta^{r}_{\mu},\,\quad e^{(2)}_{\mu}=rh_{11}\delta^{1}_{\mu}, \nonumber \\ \,\quad e^{(3)}_{\mu}&=\frac{r}{2}\frac{h_{12}}{\sqrt{h_{11}}}\delta^{1}_{\mu}+\frac{r}{2}\sqrt{ \frac{4 h_{11}h_{22} -h_{12}^2}{ h_{11} } }\delta^{2}_{\mu},  \nonumber \\
 e^{(4)}_{\mu}&=\frac{r}{2}\frac{h_{13}}{\sqrt{h_{11}}}\delta^{1}_{\mu}+ 
\frac{r}{2}  \frac{2 h_{23}h_{11} -h_{12}h_{13}}{ \sqrt{h_{11}(4 h_{11}h_{22} -h_{12}^2) }}  \delta^{2}_{\mu} \nonumber \\{}& + r\sqrt{ \frac{h_{11}( h_{23}^2-4h_{22}h_{33})+h_{12}(  h_{12}h_{33}-h_{13}h_{23})  +h_{13}^2 h_{22} }{ h_{12}^2-4h_{11}h_{22} } }\delta^{3}_{\mu}.
\end{align}

The other main ingredient is the energy-momentum tensor (\ref{eq:TmunuNLE1}), which, upon evaluation on-shell,  acquires the simple block diagonal form
\begin{align}
\label{eq:T_onshell}
 T_{\mu\nu}dx^\mu dx^\nu= \frac{ \alpha \lambda^2}{\kappa r^2} \left(-\frac{   f(\sqrt{K+6\mu}-6r\lambda^2)    }{r} dt^2+\frac{   \sqrt{K+6\mu}-6r\lambda^2   }{rf} dr^2 - 2\lambda^2  h_{ij}dx^i dx^j \right),
\end{align}
where $f$ is understood as in eq. \eqref{eq:solnf}. As advertised before, the proper classification involves moving the energy-momentum \eqref{eq:T_onshell} to the orthonormal frame via the relation $T_{AB}:=e_{(A)}^{\mu}e_{(B)}^{\nu}T_{\mu\nu}$. In doing so, the energy-momentum tensor becomes fully diagonal with components
\begin{align}\label{eq:TAB}
(T_{AB})=\frac{\alpha \lambda^2}{\kappa r^2} \left( K+6\mu\right)  \text{diag}\left(  -\frac{   \sqrt{K+6\mu}-6r\lambda^2    }{r}, \frac{   \sqrt{K+6\mu}-6r\lambda^2   }{r},  -2\lambda^2, -2\lambda^2, -2\lambda^2 \right).
\end{align} 
It is straightforward to analyze the eigenvalue problem of $T_{AB}$. A quick inspection reveals that, in the vielbein frame, a one-time and four-space-like eigenvector exists. These correspond to a non-degenerate negative eigenvalue, a non-degenerate positive eigenvalue, and three degenerate negative eigenvalues. A tensor with this algebraic structure is classified as Segre type $I$, following Ref. \cite{Maeda:2018hqu}. Under this approach, one might recognize effective energy density ($\rho$) and pressures ($p_a$, with $a=1,2,3,4$)  --- in analogy with a perfect fluid--- according to
\begin{align}
\rho:=T_{00}, \,\quad p_{1}:=T_{11}, \,\quad  p_{2}:=T_{22}, \,\quad p_{3}:=T_{33}, \,\quad p_{4}:=T_{44}.
\end{align}
Notice from the orthonormal components \eqref{eq:TAB} that, in our configuration, $\rho=-p_1$ and  $p_2=p_3=p_4$. In this form, we can write down expressions for any kind of energy conditions. We summarize in Tab.~\ref{Tab:EC} the four standard conditions that are pertinent for a wide variety of phenomenological applications.  
\begin{table}[H]\centering
\begin{tabular}{@{} *2l @{}}    \toprule
\emph{Energy condition}  & \emph{Necessary and sufficient conditions}  \\ \midrule
Strong energy condition (SEC) & $  \rho+p_{a} \geq 0, \,\,\,\forall p_a, \,\quad \text{and} \, \quad (D-3)\rho+ \sum_{a=1}^{D-1}p_{a} \geq 0 $   \\
Dominant energy condition (DEC) &  $\rho\geq 0 ,  \,\quad \text{and} \,\quad \rho+ | p_{a} | \geq 0, \,\,\,\forall p_a $ \\
 Weak energy condition \,\,(WEC) & $  \rho\geq 0,  \,\quad \text{and} \, \quad  \rho+p_{a} \geq 0, \,\,\,\forall p_a $  \\
Null energy condition \,\,\,\, (NEC) & $   \rho+p_{a} \geq 0, \,\,\,\forall p_a $   \\
 \bottomrule
 \hline
\end{tabular}
 \caption{\label{Tab:EC} Invariant energy conditions as given in terms of the energy density and effective pressures \cite{Myers:1986un}. }
 \end{table}
 It can be readily verified from the effective conditions presented in Tab.~\ref{Tab:EC} that if either the WEC or the SEC are met, then the NEC is satisfied in consequence. At the same time, if the DEC is imposed, the WEC is automatically guaranteed. This hierarchy of the energy conditions holds beyond the energy-pressure representation above. In general, the SEC and DEC are the most stringent energy conditions, while the NEC is the most permissive. In our particular scenario, one finds through evaluation of the inequalities in Tab.~\ref{Tab:EC} that the SEC is indeed very restrictive and can only be verified via a fine-tuning of the parameters $\mu$, $\lambda$, and the spatial topology. The condition responsible for this fact reads
 $$ (D-3)\rho+ \sum_{a=1}^{D-1}p_{a} =-\frac{\alpha \lambda^2}{r^3} \left(K+6\mu \right)^{3/2}\geq 0, $$
where, due to the positive definiteness of $\alpha$, $r$ and $\lambda^2$, the only alternative is the saturation of the inequality. In this regard, we encounter two options 
\begin{align} \label{eq:SEC_KM}
 K + 6\mu =0  \qquad \text{or}\qquad \lambda=0.
\end{align} 
Both conditions have a drastic consequence on the configuration. As elaborated in Sec.~\ref{sec:Thermo}, imposing the condition \eqref{eq:SEC_KM} would mean that the global charge \eqref{eq:term} is set to zero. Analogously, the solutions presented here would satisfy these conditions only if the electrodynamics is turned off, revealing thus that the only black holes admitting SECs are the uncharged family reported in \cite{Dotti:2007az}.
\\

When it comes to the DEC, WEC, and NEC, the restrictions involved are summed up in only two inequalities valid for the three  types of conditions
  \begin{align}  
 K+6\mu \geq 0, \qquad \text{and}\qquad 4\lambda^2r - \sqrt{K+6\mu} \geq 0.
 \end{align}
 Although the second expression depends on local coordinates, global constraints can still be imposed thanks to the domain where the coordinate $r$ is defined. The trick resides in exploring the two opposing limits available: the asymptotic and the near-horizon limit. The first, $r\rightarrow +\infty $, renders the condition true for any choice of constants that ensures the radical in the second inequality remains real. Conversely,  $r=r_h$ defines the other physically meaningful boundary, where the first term takes the smallest value outside the horizon.  Therefore, setting the stronger restrictions
\begin{align} \label{eq:WNEC_constraints}
 K+6\mu \geq 0, \qquad  \text{and}\qquad 4 \lambda^2r_{h} - \sqrt{K+6\mu} \geq 0,
 \end{align}
ensure that the WEC and NEC are satisfied throughout the exterior region of the BH $r\geq r_h$. Here, $r_h$ is understood as in eq.~\eqref{eq:rh}, which means that the second condition is actually an inequality involving all the constants of the solution. These conditions are translated into less rigid bounds as compared to that of the SEC, which can be enforced on our configuration with the effect of restricting the moduli space to a  physically acceptable one. Concretely, through the inequality (\ref{eq:WNEC_constraints}) together with the explicit form of $r_h$ (\ref{eq:rh}), we arrive to the parameter constraint
\begin{align*}
8\alpha\lambda^4\sqrt{K+\left(6+\frac{1}{\alpha\lambda^4}\right)\mu  } -\left( 1+8\alpha\lambda^4\right)\sqrt{K+6\mu}\geq0,
 \end{align*}
which can be written in simple terms as
 \begin{align}\label{eq:NEC}
     K + \left(6 + \frac{1}{\alpha \lambda^4}\right)\mu \ge 0,\qquad (1 + 16 \alpha \lambda^4)K \le - (6 + 32 \alpha \lambda^4)\mu\,.
 \end{align}

\begin{figure}[t]
  \captionsetup{justification=justified,singlelinecheck=false}
  \centering

  \begin{minipage}[t]{0.48\textwidth}
    \centering
    \includegraphics[width=\linewidth]{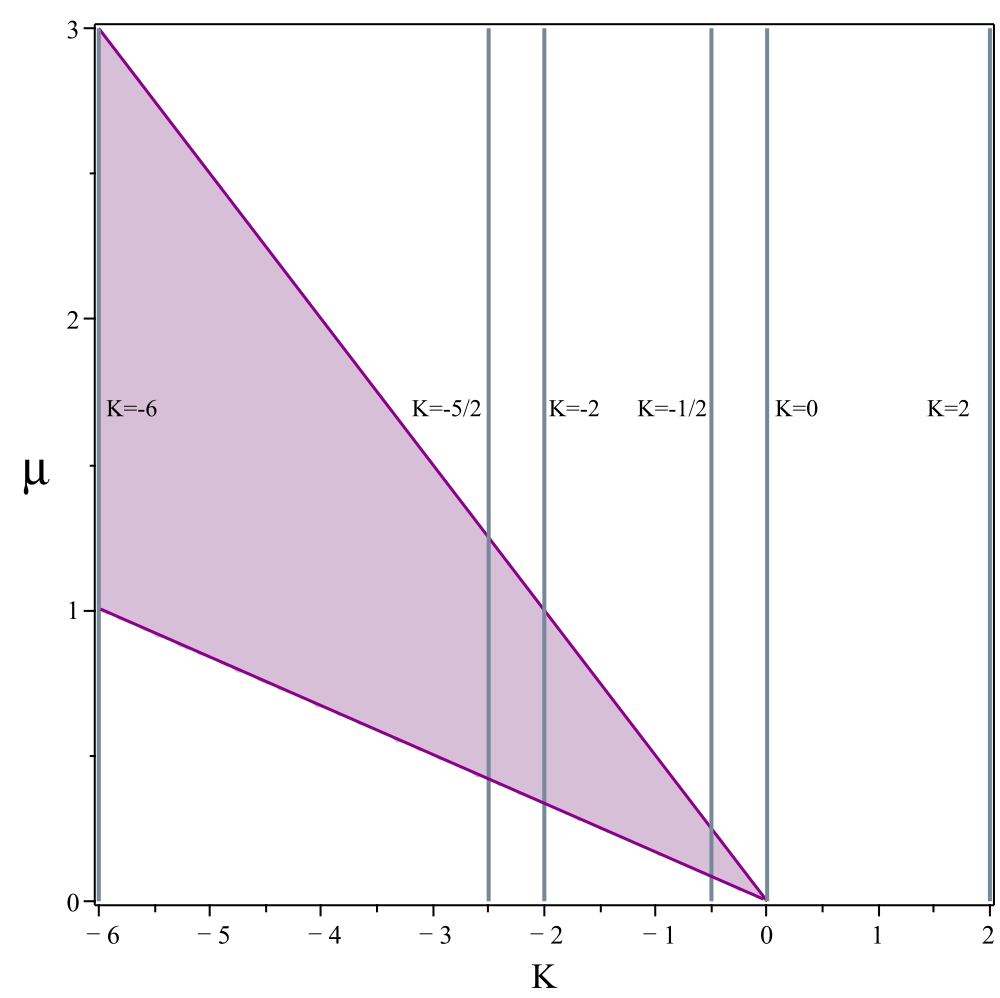}
    \par\smallskip\small (a) $\alpha \lambda^4=1000$
  \end{minipage}\hfill
  \begin{minipage}[t]{0.48\textwidth}
    \centering
    \includegraphics[width=\linewidth]{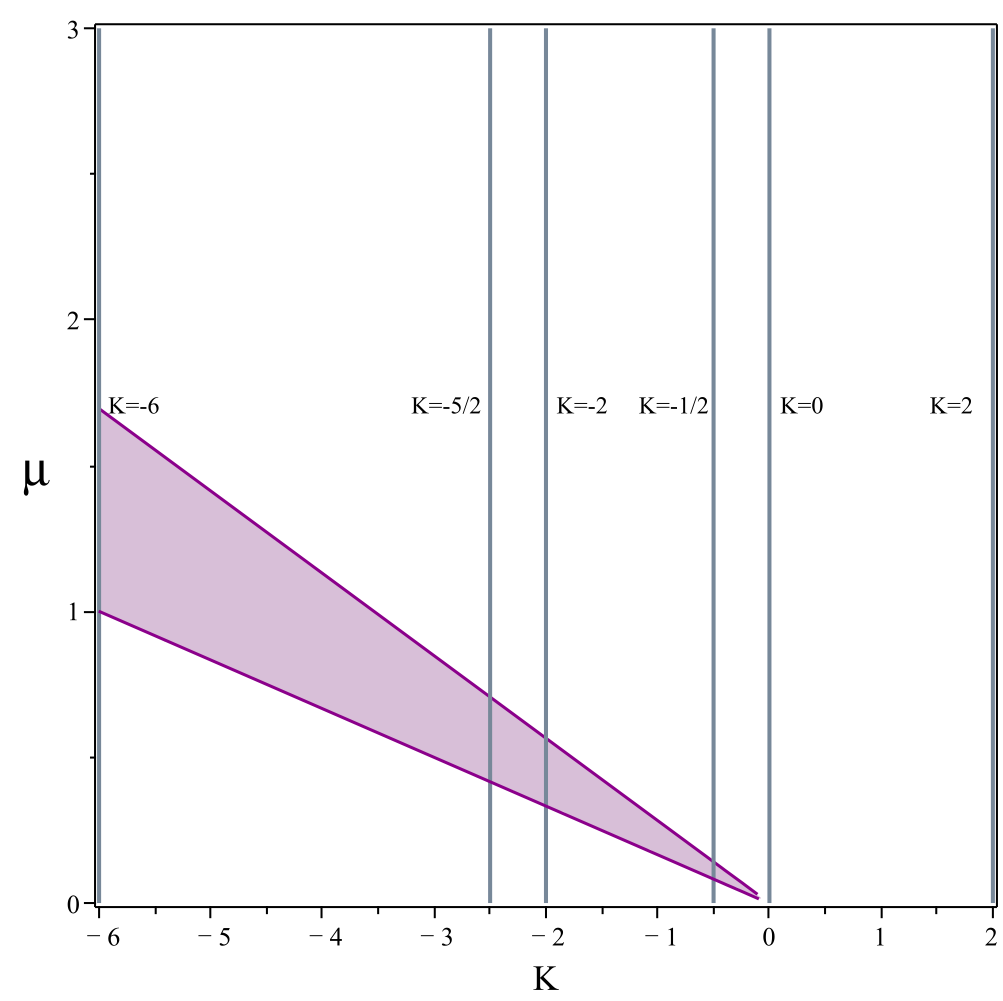}
    \par\smallskip\small (b) $\alpha \lambda^4=0.1$
  \end{minipage}

  \caption{The magenta region in the above plots displays the allowed parameter region where the WEC and NEC \eqref{eq:NEC} are satisfied. The bordering darker lines in the triangular regions are admissible points. The vertical blue lines are constant $K$ regions (see Tab.~\ref{tab:Thurston}).}
  \label{fig:WNEC}
\end{figure}
\noindent\\
It is worth pointing out that in this relationship, the Gauss-Bonnet coupling and the parameter $\lambda$ always appear together as $\alpha \lambda^4$, resulting in an inequality that effectively depends on three variables. A graphical inspection of the parameter spaces comes in handy. Fig.~\ref{fig:WNEC} shows the regions on the $K-\mu$ plane where \eqref{eq:NEC} is valid, for fixed values of $\alpha \lambda^4$. 
 
The outcome is a triangular region bounded by two lines that converge at the origin, where the inequality is saturated. Recall that along the $K$ axis, we are interested in the values of spatial curvature to each Thurston curvature in Tab.~\ref{tab:Thurston}. In that sense, we may confirm that no spacetime with $K>0$ satisfies either of the energy conditions, regardless of the value of $\alpha\lambda^4$.

Summing up, our novel BH solutions comply with the WEC and NEC, excluding only those families with $\mathbb{S}^3$ and $\mathbb{S}^1\times\mathbb{S}^2$ horizon topology. It is worth noticing that this is true for any parameter combination $\alpha\lambda^4>0$. The effect of varying such value results in narrowing the valid region while letting the bottom boundary line be fixed, as can be appreciated in Fig.~\ref{fig:WNEC}. The only problematic case is either the uncharged case ($\lambda=0$) or when the GB coupling goes to zero ($\alpha=0$), cases that are already excluded when deriving the solution. 

\section{Thermodynamics\label{sec:Thermo} }
 
In \cite{Gibbons:1976ue}, Gibbons and Hawking showed that the partition function of a thermodynamic ensemble can be identified with the Euclidean path integral, using the saddle point approximation around the analytic continuation of the solution. The time coordinate is now periodic,  with its period denoted as $\beta$, which is identified as the inverse of the temperature, $\beta = T^{-1}$. Thus, the Euclidean action is related to the Gibbs free energy $\mathcal{F}$ through 

\begin{equation}\label{eq:Gibbsfree}
S_{E}= \beta \mathcal{F} = \beta \mathcal{M} - \mathcal{S} - \beta \Phi_e\mathcal{Q}_e,
\end{equation}
where $\mathcal{M}$ stands for the mass, $\mathcal{S}$ is the entropy, $\Phi_e$ is the electric potential, and $\mathcal{Q}_e$ is the electric charge. To obtain the Gibbs free energy, it is enough to work over the following class of static Euclidean metrics:
\begin{equation}
\label{metrictermo}
ds^2 = N^2(r)f(r)d\tau^2 + \dfrac{dr^2}{f(r)} + r^2d\Sigma_3^2,
\end{equation}
where {$\tau \in [0,\beta], r\geq r_{h}$} and, as before, we will restrict $\Sigma_3$ to be a compact and orientable 3-manifold with constant Ricci scalar $K$ {with a finite volume $\displaystyle |\Omega_3|\equiv \iiint_{\Sigma_3}d\Sigma_3$.} {Then, evaluating (\ref{eq:action}) on (\ref{metrictermo}), and assuming that $A_{\mu} dx^{\mu} = A_{\tau}(r)~d\tau$ and $P_{\mu \nu} dx^{\mu} \wedge dx^{\nu}=P_{\tau r}(r) d\tau \wedge dr$, the Euclidean action can be cast in the following Hamiltonian form:
\begin{equation}\label{eq:Eaction}
S_E = \beta|\Omega_3| \int_{r_{h}}^{\infty} dr \Big( N\mathcal{H} + A_{\tau}p' \Big) + B,
\end{equation}
where
\begin{equation}
\mathcal{H}=\frac{1}{2\kappa} \left[(12 f \alpha-2K \alpha-3r^2)f'+r(-2\Lambda r^2-6 f+K)\right]+r^3 H.\label{eq:H}
\end{equation}
Here, the conjugate momentum is defined from (\ref{eq:NLE}) as
$$p \equiv -\dfrac{\delta {\cal{L}}_{\mbox{\tiny{NLE}}}}{\delta (A_{\tau}')}=\frac{r^{3} P_{\tau r} }{N},$$ and $B$ is a boundary term that matches the Euclidean action when the field equations are satisfied. Specifically, from our computations, 
\begin{equation}\label{eq:P}
P=\frac{P^2_{\tau r}}{2 N^2} =\frac{p^2}{2 r^6}.
\end{equation}
Further, to obtain the Gibbs free energy, one demands in the first place that the Euclidean action possess an extremum around the solution, i.e. $\delta S_E=0$  \cite{Regge:1974zd}. This permits us to compute the variation of the boundary term, using (\ref{eq:Eaction}). With all the above ingredients, in this case, we have that the variation
with respect to the dynamical fields $N$, $f$, $A_{\tau}$ and $p$ yield
\begin{eqnarray}
\label{eq:EN}E_{N}&:=&\frac{1}{2\kappa} \left[f'(-12 f \alpha+2 K \alpha+3 r^2)+6r f+2 r^3 \Lambda-r K\right]+r^3 H=0,\\
\label{eq:Ef}E_{f}&:=& (-12 f \alpha+2 K \alpha+3 r^2) N'=0, \\
\label{eq:EA}E_{A}&:=& p'=0, \\
\label{eq:Ep}E_{p}&:=&-H_{P} P_{p} r^3 N+A'=0,
\end{eqnarray}
where $P_p:={\partial P}/{\partial p}=p/r^6.$ Note that, from eq. (\ref{eq:EA}),  $p$ is a constant that, as we will see in the following lines, is related to the electric charge. One may verify that the system of equations is consistent with the solution (\ref{eq:solnf})-(\ref{eq:solnFrt}) presented earlier, where, for the structural function $H$ in (\ref{eq:HP_sol}), $P$ has been replaced by $-P$.
%
%

The variation of the boundary terms from (\ref{eq:Eaction})-(\ref{eq:P}) read
\begin{equation}
\delta B =-\dfrac{\delta S_E}{\delta f'} \delta f-\dfrac{\delta S_E}{\delta p'} \delta p=\beta |\Omega_3| \left[-\frac{N}{2\kappa} (12 f \alpha-2K \alpha-3r^2) \delta f - A_{\tau}\delta p\right],\label{eq:dB}
\end{equation}
where, to obtain the explicit expression of $\delta B$, the evaluation at infinity and the location of the event horizon is required. As a first case, the variations of the fields are given by
\begin{eqnarray}\label{eq:deltafp}
\delta f = \left(\frac{3\lambda^2r}{\sqrt{K+6\mu}}-1\right)~ \delta \mu,\quad p=(K+6\mu)^{3/2}\lambda \Rightarrow \delta p = 9\lambda \sqrt{K+6\mu} \,\delta \mu,
\end{eqnarray}
so that, from (\ref{eq:dB}), 
$$\delta B = \frac{\beta |\Omega_3| \alpha \delta \mu}{\kappa} (18\lambda^4 r^2-9 \lambda^2 \sqrt{6\mu+K} r+6 \mu+K)+\frac{\beta |\Omega_3| \alpha \delta \mu}{\kappa}(-18\lambda^4 r^2+9 \lambda^2 \sqrt{6\mu+K} r).$$
It is interesting to observe that the first part on the right-hand side of the last equation corresponds to the contribution of the gravity theory, while the subsequent section pertains to the matter source given by the non-linear electrodynamics. Notably, there is an exact cancellation of the linear and quadratic parts between the two terms, enabling the derivation of a finite quantity as follows:
$$\delta B(\infty) \equiv \lim_{r \rightarrow +\infty} \delta B(r) =\frac{\beta|\Omega_3| \alpha(6\mu+K) \delta \mu}{\kappa}.$$
On the horizon, considering
\begin{eqnarray}
\delta f \big{|}_{r=r_h} = -\dfrac{4\pi}{\beta}\delta r_{h},
\end{eqnarray}
the variation of the boundary term results in
$$\delta B\big{|}_{r=r_h}= \frac{2 |\Omega_3|  (2K \alpha+3 r_h^2) \pi \delta r_h}{\kappa}-\beta |\Omega_3|  A(r_h) \delta p.$$ 
Working on the grand canonical ensemble in such a way that $\beta$ and $\Phi_e$ are fixed; this term can be directly integrated to be $$ B \equiv B(\infty) - B(r_h) = \frac{\beta|\Omega_3| \mu \alpha(3\mu+K)}{\kappa}  - \frac{2 |\Omega_3|  (2K \alpha+ r_h^2) \pi r_h}{\kappa} - \beta   \Phi_e p,$$
where the Hawking temperature reads
\begin{eqnarray}\label{eq:T}
T =\frac{1}{\beta}=\frac{f'}{4 \pi}\Big{|}_{r=r_h}=\frac{r_h+2\lambda^2 \alpha \sqrt{6\mu+K}}{8\alpha\pi},
\end{eqnarray}
and the electric potential is defined by
$$\Phi_e\equiv -A_{\tau}(r_{h})=-{\frac {2{\lambda}^{3}\alpha\,r_h^{2}}{\sqrt {K+6\,\mu}\kappa
}}+{\frac {\lambda\,\alpha\,r_{h}}{\kappa}}
,$$ 
with $p$ given previously in (\ref{eq:deltafp}). Altogether, and comparing with the Gibbs free energy (\ref{eq:Gibbsfree}), the thermodynamic quantities take the form
\begin{eqnarray}
\mathcal{M}&=&\frac{|\Omega_3| \mu \alpha(3\mu+K)}{\kappa},\\
\mathcal{S}&=&\frac{2 |\Omega_3| \pi r_h(2 K \alpha+r_h^2)}{\kappa} ,\\
 \Phi_e&=&-{\frac {2{\lambda}^{3}\alpha\,r_h^{2}}{\sqrt {K+6\mu}\,\kappa
}}+{\frac {\lambda\,\alpha\,r_{h}}{\kappa}} ,\\
\mathcal{Q}_e&=&(K+6\mu)^{3/2}\lambda |\Omega_3|.\label{eq:term}
\end{eqnarray}
}
\noindent
For the sake of completeness, the thermodynamic quantity ${\cal Q}_e$ defined previously corresponds to the conserved electric charge. Although it is proportional to the parameter $\lambda$, the latter remains a coupling constant of the nonlinear electrodynamics and should not be identified with the physical charge. Additionally, we note that
\begin{align}\label{eq:TdsqP}
T \delta \mathcal{S}+\Phi_e \delta \mathcal{Q}_e= \,&{\frac { \left( r_h+2\,\lambda^{2} \alpha\sqrt {6\,\mu+K}
 \right) \left( 2\,K\alpha+3\,r_h^{2} \right) |\Omega_3| \delta r_h}{4 \kappa\,\alpha}} \nonumber\\
&+\frac {9 {\lambda}^{2}\alpha\,r_h\,
 \left( -2\,\lambda^{2}r_h+\sqrt {6\,\mu+K} \right) |\Omega_3|\delta \mu}{\kappa},
\end{align}
where, from the metric function $f$ in eq. (\ref{eq:solnf}), we must consider the relation
$$\frac{r_h^{2}}{4 \alpha}= -{\lambda}^{2}\sqrt {6\,\mu+K} r_h+\mu ,
$$
which translates into a functional dependence between $\delta r_h$ and $\delta \mu $
$$ \left({\frac {{r_h}}{2 \alpha}}+{\lambda}^{2}\sqrt {6\,\mu+K}
 \right) {\delta r_h}+ \left( {\frac {3 \lambda^{2}{r_h}}{\sqrt {6
\,\mu+K}}}-1 \right) {\delta \mu}=0.
$$
Finally, with the above expressions,  (\ref{eq:TdsqP}) turns into the well-known form
\begin{equation}
T \delta \mathcal{S}+\Phi_e \delta \mathcal{Q}_e=\frac{|\Omega_3|\alpha (6\mu+K) \delta \mu}{\kappa}=\delta \mathcal{M},\label{eq:first-law}
\end{equation}
ensuring the first law of black hole thermodynamics.

In addition to the previous discussion, several comments can be made regarding the non-negativity of the thermodynamic quantities presented earlier in (\ref{eq:T})-(\ref{eq:term}). The first observation is that, from the expression of the metric function $f$ in eq. (\ref{eq:solnf}), and to ensure an active contribution from its linear part and a well-defined Maxwell tensor (\ref{eq:solnFrt}), we find that $K+6\mu>0$. Furthermore, we obtain $\mathcal{Q}_e\geq 0$ and $T \geq 0$ for $\lambda \geq 0$. Additional restrictions on the positive integration constant $\mu$ must be considered. In particular, to achieve $\mathcal{M}\geq 0$, ${\Phi}_e \geq 0$, and $\mathcal{S} \geq 0$, the constraint $r_h \geq 0$ should be supplemented by
\begin{eqnarray} \label{eq:in1}
3\mu+K \geq 0, \qquad -2\lambda^2r_h+\sqrt{K+6\mu} \geq 0,\qquad 2 K\alpha+r_h^2 \geq 0,
\end{eqnarray}
respectively, where together with the positive branch from eq. (\ref{eq:rh}), the last two inequalities of (\ref{eq:in1}) become to
\begin{eqnarray}
 (32\lambda^4\alpha+6)\mu+K (8\lambda^4\alpha+1)\geq 0, \label{eq:in2}\\
\mu^2+K(12\lambda^4\alpha+1)\mu+\frac{1}{4} K^2 (8\lambda^4\alpha+1) \geq 0. \label{eq:in3}
\end{eqnarray}
Obviously, from Table \ref{tab:Thurston} for $K \geq 0$, we have that all the thermodynamic quantities are non-negative. Nevertheless, for $K<0$ {\em{a priori}} an additional condition for the integration constant $\mu$ must be considered.  The first condition, to ensure $\mathcal{M} \geq 0$, requires imposing $\mu \geq -K/3$ from (\ref{eq:in1}). It's worth noting that the above inequality allows us to satisfy (\ref{eq:in2}), while the second one (\ref{eq:in3}) is fulfilled only if $1-72 \alpha \lambda^4 \geq 0$.

\begin{figure}[H]
  \centering
 \includegraphics[scale=0.28]{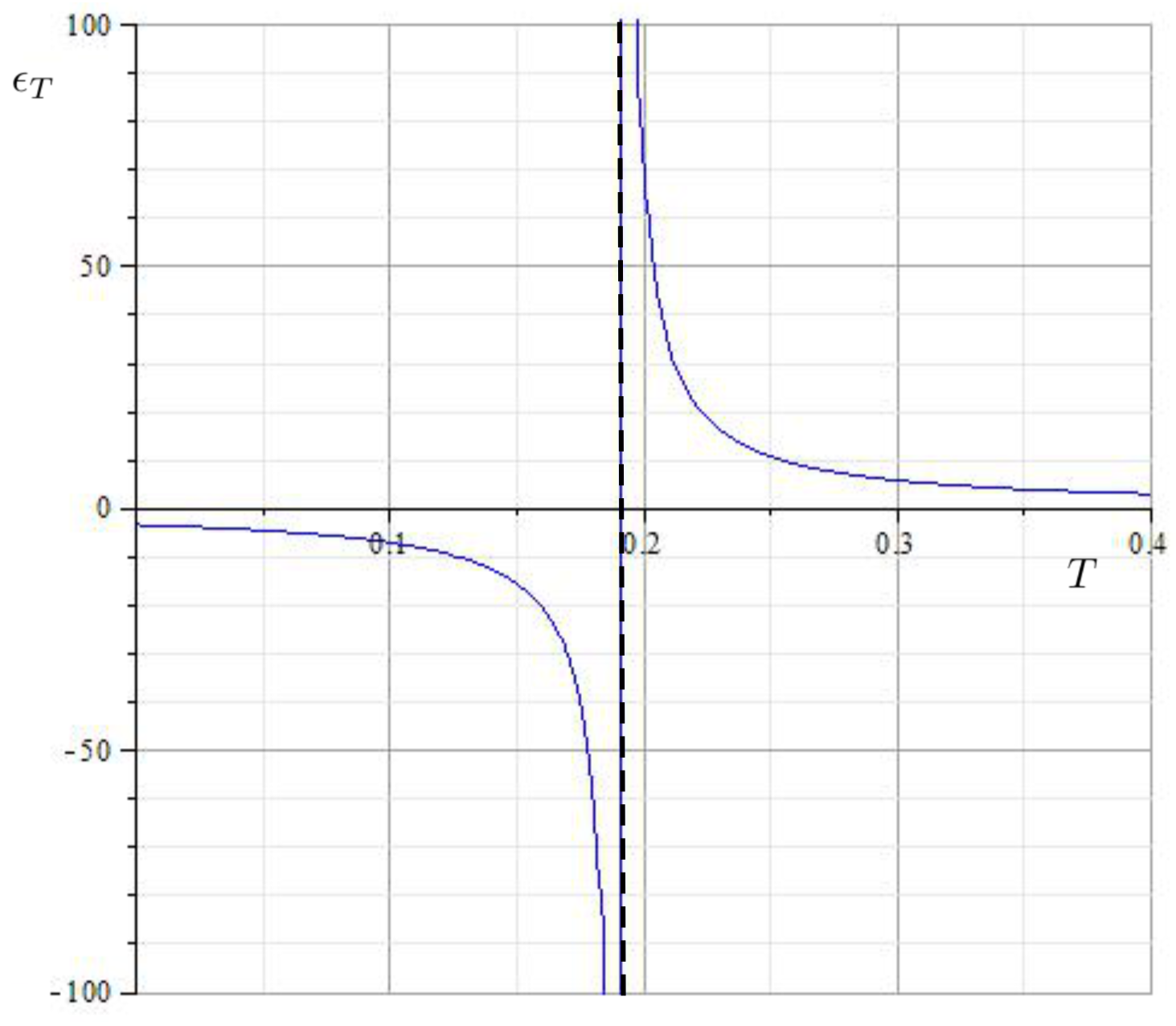}
      \caption{\label{fig:et} 
      Electric permittivity as a function temperature $T$, for $\Phi_e=0.4$ and $\mathcal{Q}_{e}=1$. Positive and negative branches are present in this case, along with a vertical asymptote corresponding to $\Psi=0$. Here, we have considered $\lambda=4 \alpha=|\Omega_{3}|=\kappa=1$ for our computations. }
 
\end{figure}

On the other hand, additional insight can be gained by expressing the location of the event horizon in terms of the charge and temperature using eqs. (\ref{eq:T}) and (\ref{eq:term})
\begin{equation}\label{eq:rh2}
r_h=\frac{2 \alpha( 4 T  |\Omega_3|^{\frac{1}{3}}\pi-\lambda^{\frac{5}{3}}\mathcal{Q}_e^{\frac{1}{3}})}{|\Omega_3|^{\frac{1}{3}}},
\end{equation}
implying that $4 T |\Omega_3|^{\frac{1}{3}}\pi - \lambda^{\frac{5}{3}}\mathcal{Q}_e^{\frac{1}{3}} \geq 0$. Eq. (\ref{eq:rh2}) is a derived thermodynamic relation between $r_h$ in terms of the temperature $T$ and the electric charge ${\cal Q}_e$. The explicit dependence on $\lambda$ reflects its role as a coupling constant of the nonlinear electrodynamics, and it cannot be absorbed into the definition of the electric charge without changing the underlying theory.  The extensive parameters $\mathcal{Q}_e$ and $\mathcal{S}$ can be expressed as functions of $T$ and $\Phi_e$ through the following implicit relations:
\begin{eqnarray}
2\,{\lambda}^{\frac{8}{3}}{\alpha}^{2} \left( 4\,\alpha\,{\lambda}^{4}+1
 \right) \mathcal{Q}_e^{\frac{2}{3}}+ |\Omega_3|^{\frac{1}{3}}\left( \Phi_{e}
\kappa-8\,T \pi \,{\alpha}^{2}\lambda\, \left( 8
\,\alpha\,{\lambda}^{4}+1 \right)  \right) \mathcal{Q}_e^{\frac{1}{3}}+128\,{
\lambda}^{\frac{10}{3}}{\alpha}^{3}{T}^{2}|\Omega_3|^{\frac{2}{3}} {\pi }^{2}&=&0,\label{eq:me1}\\
8 \pi \alpha^2 \lambda^2 (8\alpha \lambda^4+1)\mathcal{Q}_e-32\pi^2\alpha^2 T |\Omega_3|^{\frac{1}{3}}\lambda^{\frac{1}{3}}(12\alpha\lambda^4+1)\mathcal{Q}_e^{\frac{2}{3}}+\mathcal{S}\lambda \kappa+2048\pi^4\alpha^3 T^3 |\Omega_3| \lambda&=&0.\label{eq:me2}
\end{eqnarray}
\noindent
With this information, we may analyze the local thermodynamic (in)stability of this charged BH solution under electrical and thermal fluctuations, ensuring that all of the derived thermodynamic quantities are non-negative and thus satisfy the previous relations (\ref{eq:in1})-(\ref{eq:in3}). The first one involves the electric permittivity $\epsilon_{T}$, obtainable from eq. (\ref{eq:me1}), and is given by
\begin{align}
\epsilon_{T}=&\left(\frac{\partial \mathcal{Q}_e}{\partial \Phi_e}\right)_{T}=\frac{3 |\Omega_3|^{\frac{1}{3}}\kappa \mathcal{Q}_e}{8|\Omega_3|^{\frac{1}{3}} \pi T \alpha^2\lambda (8\alpha\lambda^4+1)-4\lambda^{\frac{8}{3}}\alpha^2 \mathcal{Q}_e^{\frac{1}{3}}(4\alpha\lambda^4+1)- |\Omega_3|^{\frac{1}{3}}\Phi_e\kappa} \nonumber \\
=&\frac{3 |\Omega_3|^{\frac{1}{3}}\kappa \mathcal{Q}_e}{\Psi},\label{eq:perm}
\end{align}
where the  sub-index stands for a constant temperature $T$, and $\Psi$ is defined by
\begin{equation}
    \Psi=8|\Omega_3|^{\frac{1}{3}} \pi T \alpha^2\lambda (8\alpha\lambda^4+1)-4\lambda^{\frac{8}{3}}\alpha^2 \mathcal{Q}_e^{\frac{1}{3}}(4\alpha\lambda^4+1)- |\Omega_3|^{\frac{1}{3}}\Phi_e\kappa.
\end{equation}
Here we note that if 
\begin{equation}\label{eq:ineq}
\Psi>0, \quad \mbox{ and } \quad \mathcal{Q}_{e} \geq 0,
\end{equation}
the non-negativity of the quantity $\epsilon_T$, and the subsequent local stability under electrical fluctuations, is guaranteed. It is worth pointing out that this BH is electrically unstable if the electric permittivity is negative; due to that, the electric potential decreases while the system acquires more electric charge, leaving the equilibrium state \cite{Chamblin:1999hg}. In our situation,  this criterion hinges on the sign of $\Psi$, assuming a non-negative electric charge. As depicted in Fig. \ref{fig:et}, $\epsilon_{T}$ exhibits both positive and negative branches and features a vertical asymptote. 
    
\begin{figure}[H]
  \centering
 \includegraphics[scale=0.15]{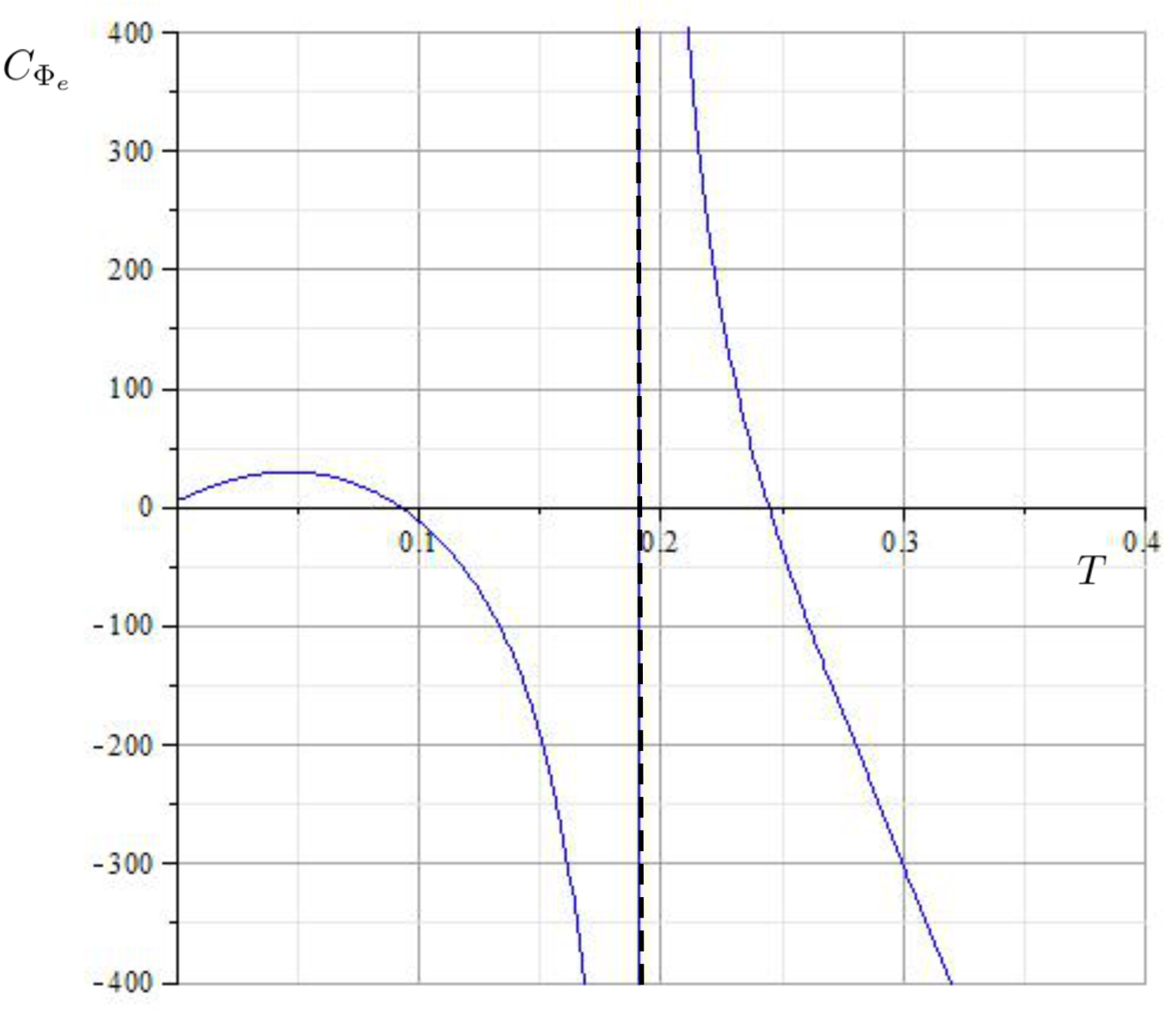}
      \caption{
      \label{fig:cphi} 
      Specific heat as a function of the temperature $T$, where $\Phi_e=0.4$ and $\mathcal{Q}_{e}=1$. In this case, positive and negative branches are observed, as well as a vertical asymptote represented by $\Psi=0$. Here, we have considered $\lambda=4 \alpha=|\Omega_{3}|=\kappa=1$ for our computations. }
      
\end{figure}

On a different note,  the local thermodynamic stability under thermal fluctuations can be determined through
the behavior of the specific heat $C_{\Phi_{e}}$, where, from eq. (\ref{eq:me2}), it takes the form 
\begin{eqnarray} \label{eq:spec_heat}
C_{\Phi_{e}}&=&T \left(\frac{\partial \mathcal{S}}{ \partial T}\right)_{\Phi_{e}}=\frac{32 \pi^2 \alpha^2  \Sigma }{\kappa \lambda \Psi },
\end{eqnarray}
with
\begin{eqnarray}
\Sigma&=& T \Big\{2\alpha^2\lambda^3 |\Omega_3|^{\frac{1}{3}} (96\lambda^8 \alpha^2+16 \alpha \lambda^4+1)  \mathcal{Q}_e-|\Omega_3|^{\frac{2}{3}} \big[8 \pi \lambda^{\frac{4}{3}} T \alpha^2 (8\alpha\lambda^4+1) (36\alpha \lambda^4+1)\nonumber\\
&+&\lambda^{\frac{1}{3}}\Phi_e\kappa (12\alpha \lambda^4+1)\big] \mathcal{Q}_e^{\frac{2}{3}}+256\pi^2\alpha^3|\Omega_3| T^2\lambda^{\frac{11}{3}} (5+36\alpha \lambda^4) \mathcal{Q}_e^{\frac{1}{3}} \nonumber\\
&-&1536 \pi^3|\Omega_3|^{\frac{4}{3}} T^3 \alpha^3 \lambda^2 (8\alpha\lambda^4+1)+192 T^2 \lambda \pi^2 \alpha |\Omega_3|^{\frac{4}{3}}\Phi_e \kappa\Big\},\label{eq:Sigma}
\end{eqnarray}
where, analogously, the sub-index stands for a constant electrical potential $\Phi_{e}$. Local stability under thermal fluctuation is guaranteed when $\Sigma \geq 0$ and $\Psi>0$, with $\lambda>0$. Similar to the electric permittivity case, the local stability under thermal fluctuation depends on the sign of $\Sigma$ and $\Psi$, where $\Psi=0$ once again represents vertical asymptote, as shown in Fig. \ref{fig:cphi}.

It is notorious that satisfying (\ref{eq:ineq}) alongside $\Sigma \geq 0$ (with $\lambda>0$) renders this charged configuration locally stable under both thermal and electrical fluctuations. 

\section{Transport coefficients \label{sec:Transport} }
 


As previously stated, we will follow the approach outlined in reference \cite{Arias:2017yqj}, which leverages the non-shift invariant nature of the base manifold to allow for finite values of conductivities in electrically charged BH configurations. Through the metric and gauge field perturbations, thermoelectric DC conductivities can be derived from the horizon data \cite{Donos:2014cya}. This method involves constructing conserved currents from the linearized equations, considering the corresponding boundary conditions.
Such a scheme has also been explored with non-standard Maxwell sources; see for instance, references \cite{Bi:2021maw,Rai:2019iwf,Mu:2017usw,Wang:2018hwg,Sadeghi:2021qou}. 
In our case, in addition to the unique non-linear Maxwell term \eqref{eq:HP_sol}, we have five (non-maximally symmetric) viable geometries to model the BH horizon. 
The maximally symmetric cases are excluded because, at the CS point,
the linearized Einstein-GB equations on constant–curvature backgrounds are degenerate and
do not support a well-defined linear response. Consequently, perturbative (Kubo) computations around \(\mathbb{E}^3,\mathbb{S}^3,\mathbb{H}^3\) are ill-defined and will not be pursued in what follows.

Referring to the descriptions given in Table \ref{tab:Thurston}, we will utilize the above information to comprehensively develop the construction for the $\mbox{SL}_2(\mathbb{R})$ case. In addition, we provide a few comments concerning the other geometries.

Starting from the $\mbox{SL}_2(\mathbb{R})$ metric (\ref{Sl2}), it is clear that the $x_1$ direction breaks the translation invariance. Accordingly, we perturb the gauge field along that direction as $$\delta A = (-Et + \delta a(r))~dx_1,$$ 
where $E$ coincides with the applied electric field. 
Consistency in the linear regime imposes the following background perturbations:
 $$\delta g_{tx_1} = h_{tx_1},\ \delta g_{rx_1} = h_{rx_1}.$$
In the first-order expansion, the Maxwell equations provide the following conserved current:
\begin{equation}
\label{Jmaxwell}
 J = - \left(H_P\big{|}_{P=P(r)}\right)^{-1}\left( f(r)r \delta a' + A_t' h_{tx_1} r \right),
\end{equation}
which is to be evaluated at the horizon $r=r_h$.
Moving to Eddington-Finkelstein coordinates reveals that it is sufficient to impose
\begin{equation}
\label{regcond}
\delta a' = - \dfrac{E}{f(r)} - \dfrac{4E}{3},\quad \delta g_{tx_1} = f(r)\delta g_{rx_1},
\end{equation} 
to ensure regular boundary conditions at the horizon. 
The addition of the last term in $\delta a'$ does not affect regularity; however, it is important to satisfy the linear Einstein equations, together with $$h_{tx_1} = \dfrac{\kappa(K+6\mu)E}{3\lambda \alpha}.$$
Hence, the current $J$ in eq. (\ref{Jmaxwell}) now reads
$$J = \dfrac{1}{3} \dfrac{ \kappa(K+6\mu)^{5/2}(2r^2 + 4\mu\alpha + \alpha)E }{r^2 \alpha^2 (5+8\lambda^2\sqrt{K+6\mu}r - 12\mu)},$$
and therefore, the DC conductivity $\sigma$ is
\begin{equation}
\label{sigma}
\sigma = \left. \dfrac{\partial J}{\partial E}\right|_{r=r_h} = \dfrac{1}{3} \dfrac{ \kappa(K+6\mu)^{5/2}(2r_h^2 + 4\mu\alpha + \alpha) }{r_h^2 \alpha^2 (5+8\lambda^2\sqrt{K+6\mu}r_h - 12\mu)}.
\end{equation}

The strategy to tackle down the conductivity is the same in the four remaining geometries. Nevertheless, some subtleties are worth mentioning.  For Nil and Solv geometries, the perturbation structures mirror those in \cite{Arias:2017yqj}.  Yet, in the Solv geometry (\ref{Solvg}) we need to perform the diffeomorphism $z\mapsto \ln (x_3)$ that maps the metric tensor presented in \cite{Arias:2017yqj} to the one in our manuscript. 
This induces a modification of the perturbations, which now depend on the $x_3$-coordinate as
$$ \delta g_{rx_3} = \dfrac{h_{rx_3}(r)}{x_3},\ \delta g_{tx_3}=\dfrac{h_{tx_3}(r)}{x_3},\ \delta A = (-Et + \delta a(r))~\dfrac{dx_3}{x_3}.$$ 
To ensure finite conductivities in the remaining geometries, namely $\mathbb{S}^{1}\times \mathbb{S}^2$ and $\mathbb{S}^{1}\times \mathbb{H}^2$, we found it compulsory to introduce a spatial dependence in the perturbations of the metric tensor, which can be expressed as follows: 
$$ \delta g_{rx_i} = h_{rx_i}(r)p(x_1,x_2,x_3),\ \delta g_{tx_i}=h_{tx_i}(r)p(x_1,x_2,x_3).$$ 
Similarly for the gauge field,
 $\delta A = p(x_1,x_2,x_3)(-Et + \delta a(r))~dx_i$, where $x_i$ denotes the direction in which shift-invariance is broken. 
The regularity conditions at the horizon (\ref{regcond}) manifest as $$\delta a' = -\dfrac{E}{f(r)} + \theta E,\quad \delta g_{tx_i} = f(r)\delta g_{rx_i},$$ where $\theta$ is a constant depending on the geometry. We summarize all the pertinent information in Tab. \ref{Tab:Cond}:
\begin{table}[H]\centering
\begin{tabular}{ccccc}   \toprule
\emph{Geometry}  & $x_i$ & $p(x_1,x_2,x_3)$ & $\theta$ & \emph{DC Conductivity ($\sigma$)} \\ \midrule
\vspace*{0.2cm}
\vspace*{0.2cm}
$\mathbb{S}^{1}\times \mathbb{S}^2$ & $x_2$ & $\dfrac{1}{\sin(x_2)}$ & $2$ & $\dfrac{1}{4} \dfrac{ \kappa(K+6\mu)^{5/2}(r_h^2 + 2\mu\alpha ) }{r_h^2 \alpha^2 (1-2\lambda^2\sqrt{K+6\mu}r_h + 3\mu)}$ \\
\vspace*{0.2cm}
$\mathbb{S}^{1} \times \mathbb{H}^{2}$  & $x_2$ & $\dfrac{1}{\sinh(x_2)}$ & $-2$ & $\dfrac{1}{4} \dfrac{ \kappa(K+6\mu)^{5/2}(r_h^2 + 2\mu\alpha ) }{r_h^2 \alpha^2 (1+2\lambda^2\sqrt{K+6\mu}r_h - 3\mu)}$    \\
\vspace*{0.2cm}
$\mbox{Solv}$  & $x_3$ & $\dfrac{1}{x_3}$ & $-1$ & $\dfrac{1}{8} \dfrac{ \kappa(K+6\mu)^{5/2}(r_h^2 + 2\mu\alpha + 2\alpha) }{r_h^2 \alpha^2 (1+2\lambda^2\sqrt{K+6\mu}r_h - 3\mu)}$    \\
\vspace*{0.2cm}
$\mbox{Nil}$  & $x_1$ & $1$ & $-4$ &  $\dfrac{ \kappa(K+6\mu)^{5/2}(2r_h^2 + 4\mu\alpha + \alpha) }{r_h^2 \alpha^2 (1+8\lambda^2\sqrt{K+6\mu}r_h - 12\mu)}$  \\
\vspace*{0.2cm}
$\mbox{SL}_2(\mathbb{R})$  & $x_1$ & $1$& $-\frac{4}{3}$ & $\dfrac{1}{3} \dfrac{ \kappa(K+6\mu)^{5/2}(2r_h^2 + 4\mu\alpha + \alpha) }{r_h^2 \alpha^2 (5+8\lambda^2\sqrt{K+6\mu}r_h - 12\mu)}$ \\
 \bottomrule
 \hline
\end{tabular}
 \caption{\label{Tab:Cond} Essential considerations for achieving finite conductivities across all the non-maximally symmetric geometries. Here, $x_i$ denotes the direction in which the geometry has a broken shift-invariance.}
\end{table}
\begin{figure}[h] 
  \captionsetup{justification=justified,singlelinecheck=false}
  \centering

  \begin{minipage}[t]{0.48\textwidth}
    \centering
    \includegraphics[width=\linewidth]{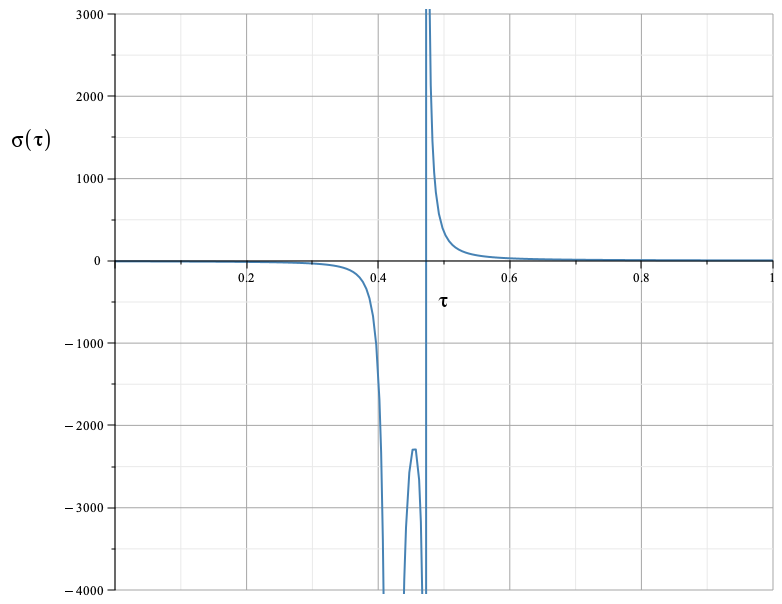} 
    \par\smallskip\small (a) $\mathbb{S}^1 \times \mathbb{H}^2$
  \end{minipage}\hfill
  \begin{minipage}[t]{0.48\textwidth}
    \centering
    \includegraphics[width=\linewidth]{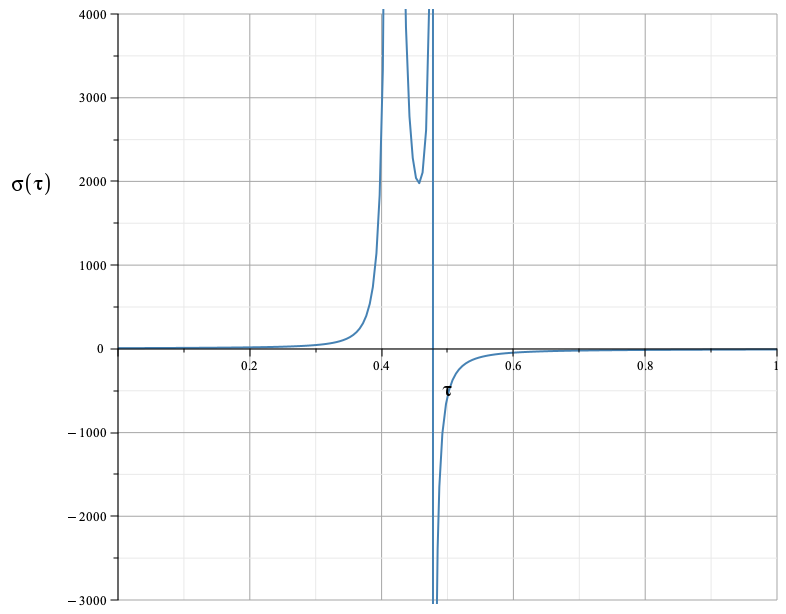}
    \par\smallskip\small (b) $\mathrm{SL}_{2}(\mathbb{R})$
  \end{minipage}

  \caption{Illustration of two characteristic behaviors of the DC conductivities as a function of the temperature. On the left, we depict the behavior stemming from a configuration with an  $\mathbb{S}^1\times \mathbb{H}^2$ horizon, while on the right, we consider the $\mbox{SL}_{2}(\mathbb{R})$ geometry. We introduce the dimensionless temperature $\tau = \lambda^2 T$, along with the test values of dimensionless quantities as $\alpha \lambda^4=1$ and $\mu=5$. This choice adheres to all energy conditions.}
  \label{fig:SigOT}
\end{figure}

To illuminate the physics underlying this result, let us first recall the  BH temperature computed in eq. \eqref{eq:T}. Subsequently, we can express the conductivity as a function of the temperature, thereby characterizing its behavior and pertinent limits.
This analytical relation $\sigma(T)$ can be derived for each geometry using the data presented in Table \ref{Tab:Cond}. Nevertheless, a more direct understanding can be attained through graphical analysis.

In Fig. \ref{fig:SigOT}, we present plots of the holographic conductivity as a function of the dimensionless temperature. Specifically, we illustrate results for BH configurations with base geometries $\mathbb{S}^1\times \mathbb{H}^2$ (Fig. \ref{fig:SigOT}a) and $\mbox{SL}_{2}(\mathbb{R})$ (Fig. \ref{fig:SigOT}b). Notably, we observe a critical temperature that delineates a negative definite branch from a positive definite one. The presence of this critical point makes evident a potential phase transition, suggesting a superconducting phase. Furthermore, as temperature increases, conductivity approaches zero, consistent with the expectations.

It is noteworthy that the remaining geometries,  Nil and Solv, exhibit a behavior similar to that of $\mathbb{S}^1\times \mathbb{H}^2$. The distinctive transition from positive to negative conductivity is observed only for the $\mbox{SL}_{2}(\mathbb{R})$ base geometry.

\section{Interpretation of the dual transport behavior\label{sec:interpretation}}

The emergence of negative DC conductivity regimes in our BH solutions, although initially unexpected, finds a meaningful parallel within the class of anomalous transport phenomena observed in non-equilibrium condensed matter systems. Historically, effects such as absolute negative conductance under strong terahertz (THz) fields~\cite{PhysRevLett.75.4102}, radiation-induced zero-resistance states in two-dimensional electron gases (2DEGs)~\cite{PhysRevLett.90.046807, PhysRevB.64.201311}, and negative magnetoresistance under nonlinear transport conditions~\cite{Bykov2005} have demonstrated that transport coefficients can undergo sign inversions when systems are driven sufficiently far from equilibrium.

In our holographic constructions, the DC conductivity $\sigma_{\text{DC}}$ exhibits temperature windows where it approaches zero (at a critical temperature $T_c$) and subsequently changes sign. The presence of zero-resistance-like points ($\sigma_{\text{DC}} \to \infty$) and inversions of the conductivity sign marks critical scales in the bulk geometry, suggesting transition-like phenomena in the boundary system. Taking into account the relation between the Hawking temperature and the horizon radius, eq.~\eqref{eq:T}, these transitions correspond to critical combinations of the horizon size and the integration constants. Accordingly, each Thurston topology supports a critical radius associated with anomalous transport behavior.

This behavior shares phenomenological features with holographic superconductors, where the vanishing of resistivity arises from the condensation of charged scalar operators and the subsequent spontaneous symmetry breaking. In contrast, our model displays similar transition-like behavior without the presence of scalar fields, suggesting that the nonlinear electrodynamics sector modulates the transport response. Although a precise classification of the observed transitions would require a fluctuation analysis, the abrupt variations in $\sigma_{\text{DC}}$ with temperature point toward structural reorganizations in the effective dynamics of the boundary theory.

Thermodynamic stability analysis performed in Sec.~\ref{sec:Thermo} further reinforces this interpretation. Both the electric permittivity $\epsilon_T$ and the heat capacity $C_{\Phi_e}$, given by eqs.~\eqref{eq:perm} and~\eqref{eq:spec_heat}, display points of discontinuity and regions of negativity, indicating local instabilities. Notably, the critical temperatures associated with these thermodynamic transitions align with those characterizing the conductivity sign changes. This correlation suggests that the anomalous transport regimes are deeply linked to the underlying thermodynamic structure of the dual system.

The sensitivity of the transport coefficients to the Thurston geometry is also noteworthy. Particularly, the $SL(2,\mathbb{R})$ horizon geometry exhibits a predominantly negative conductivity branch, separated by a characteristic temperature scale, as shown in Fig.~\ref{fig:SigOT}b. From a transport theory viewpoint, this behavior would imply a maximal temperature beyond which the dual conductor becomes unphysical. Although in our case these features emerge from the intrinsic geometry of the BH solutions, similar anomalous transport behaviors have been observed experimentally under strong external driving. For instance,~\cite{Bykov2008} reports a transition from zero-resistance to absolute negative resistance in a 2DEG system driven by microwave irradiation and magnetic fields, attributed to multiphoton-assisted impurity scattering under nonequilibrium conditions. In contrast, Nil and Solv horizon geometries, while exhibiting non-trivial temperature dependence, show milder sign-changing behavior, suggesting a subtler imprint of the spatial topology on the transport response.

Complementary to experimental observations, theoretical models in condensed matter physics predict anomalous transport regimes such as negative differential conductivity near cyclotron resonance~\cite{PhysRevB.68.115324} and absolute negative conductivity driven by photon-assisted scattering~\cite{RYZHII200413, Ryzhii2004}. However, the anomalous behavior in our solutions arises intrinsically, without external driving fields, purely from the interplay between non-linear electrodynamics and horizon topology. This intrinsic realization distinguishes the mechanism at play in the configurations presented.

Overall, the interplay between non-linear matter dynamics and the underlying Thurston geometries appears sufficient to generate rich dual transport behavior, including conductivity sign inversions and critical phenomena. While a direct microscopic identification between our holographic models and specific materials remains outside the current scope, the qualitative similarities observed ---anomalous transport, critical temperatures, and topology-dependent features--- suggest that this class of models provides a novel approach for exploring non-equilibrium phenomena from a gravitational perspective.

\section{Final remarks\label{sec:Remarks} }
In this work, we explored novel aspects of higher-dimensional Gauss-Bonnet BHs and their implications for holography.
 We have presented a new family of exact charged  BH solutions in five dimensions, extending the BTZ-like configurations originally outlined in \cite{Dotti:2007az}. 
The distinguishing feature of our geometries lies in the appearance of a linear term in $r$ inside the metric \eqref{eq:solnf}. This is due solely to the coupling of non-linear electrodynamics alongside the GB term. These configurations are, in principle, attainable for any of the eight Thurston geometries. However, as detailed in Sec. \ref{sec:BlackHole}, we determine uniquely the solutions for non-maximally symmetric base manifolds. It is left for future investigation to integrate the field equations for such cases, which necessitate more intricate scrutiny.

 Notably, the electrodynamics supporting these  BHs lack a Maxwell limit, as can be seen in eq. \eqref{eq:inv_legendre1}. It implies that non-linearity is imperative to derive a nontrivial solution. Therefore, it becomes crucial to investigate the energy conditions to assess the physicality of our solutions and the coupled fields. We prove that the standard energy conditions (DEC, WEC, and NEC) are met within a bounded parameter domain. Specifically, a series of inequalities among mass, charge, and the GB coupling must be adhered to, with the most stringent criteria imposed by the DEC and WEC.

With respect to the thermodynamics of the configurations,  the Euclidean action (\ref{eq:Eaction})-(\ref{eq:H}) yields intriguing properties. First law, for instance, is manifestly valid (see eq. (\ref{eq:first-law})). Additionally, with the information derived, we are in a position to study local thermodynamic stability under thermal and electrical fluctuations, characterized by heat capacity and electrical permittivity. Notably, we identify parameter ranges where both quantities are non-negative, permitting the existence of locally stable BHs.

In our framework, we encountered conditions that motivated the exploration of a holographic application. The presence of a higher-dimensional setup, coupled with a gauge field, and the inherent space anisotropy of the non-maximally symmetric Thurston geometries, makes it ideal for computing the holographic DC conductivities.  For this, we were able to derive analytical expressions for the conductivity in all five cases of interest, as illustrated in Tab. \ref{Tab:Cond}. Similar scenarios have been explored, both in GB and higher-derivative gravity; see, for instance,
 \cite{Cheng:2014tya, Donos:2017oym, Hao:2022zkr}. In a closely related finding, we observe that in addition to the location of the event horizon $r_h$, the conductivities depend upon the temperature,  the integration constant $\mu$, the constant $\lambda$, as well as the GB coupling constant $\alpha$. However, in the quantities we report, we also find a contribution from the constant curvature  $K$, characteristic of each base manifold.
Inspection of the analytical expressions reveals that as the temperature $T$ increases,
the conductivity $\sigma$ tends to zero.
In all five cases, we identify critical temperatures, where negative definite and positive definite branches are delineated, yielding a phase transition. 

As a final remark, akin to the indefiniteness of the geometry for the maximally symmetric cases,  non-linear electrodynamics cannot be uniquely determined. Apparently, a proper gauge fixing needs to be established to attain the full solution. Consequently, it is physically unfeasible to realize holographic models for the $\mathbb{E}^{3}$, $\mathbb{S}^3$, and $\mathbb{H}^{3}$ geometries, and obtain sensible holographic transport coefficients such as the (thermo)electric conductivities. This ultimately traces back to the dynamical inconsistency of the linear regime around the CS point, a fundamental piece missing for standard holographic applications. Extending the configurations presented here to such cases remains an open problem to be explored with higher-order methods, as it succinctly discussed in \cite{Tapia:2025vtn}.

\acknowledgments
The authors would like to thank the anonymous referee for carefully reading our manuscript and giving valuable suggestions that led to an improved version of this work.
MB is supported by PROYECTO INTERNO UCM, L\'INEA REGULAR  UCM-IN-25202.
JAMZ acknowledges funding support from a grant under the \emph{Estancias Posdoctorales por M\'exico} program from CONAHCYT.  DFHB and JAMZ thank the Sistema Nacional de Investigadoras e Investigadores (SNII) from CONAHCYT for
support.  The authors greatly appreciate the insightful commentary and recommendations provided at the outset of this work by professors Aristomenis Donos, Jerome P. Gauntlett, and Julio Oliva.

\appendix

\section{Equivalent $\mathcal{L}(F)$ formulation of the non-linear  electrodynamics \label{sec:appA}}

In Sec.~\ref{sec:BlackHole}, we have derived a family of  BHs whose configuration is reliant on the coupling of non-linear electrodynamics. The matter action, represented in eq. \eqref{eq:NLE}, models an arbitrary non-linear electrodynamics whose form, dictated by the structural function ${H}(P)$, is fixed dynamically by the field equations. However, the straightforward way to think of non-linear electrodynamics is as a generalization of the Maxwell invariant such that it is replaced in the Lagrangian by an arbitrary differentiable function of it. Namely,
   \begin{equation} \label{eq:LofF}
  		\mathcal{L}_{\text{Maxwell}}=F:=-\frac{1}{4}F_{\mu\nu}F^{\mu\nu} \qquad \mapsto \qquad \mathcal{L}_{\text{NLED}} =\mathcal{L}(F).
   \end{equation}
   While both formulations \eqref{eq:LofF} and (\ref{eq:NLE}) are known to be formally equivalent, it is not always possible to integrate explicitly one form when the other is given.  Luckily, the structural function found in \eqref{eq:HP_sol} can be expressed in the 
  $\mathcal{L}(F)$ formalism. For completeness, we proceed to construct this relation, which might be more convenient for certain applications.
  \\
 The transition from  one formulation to the other is achieved through the differential equation posed by the relation 
\begin{align} \label{eq:NLED_HP}
\mathcal{L}_{\text{NLED}}=\frac{1}{2}F_{\mu\nu}P^{\mu\nu}-{H}\left( {P} \right) = \mathcal{L}(F)  ,
\end{align}
 stating that the equivalence of the field dynamics is realized through different field variables. Here, eq.~\eqref{eq:NLED_HP} must be supplemented with the on-shell link between $F_{\mu \nu}$ and $P_{\mu \nu}$ given by
\begin{align} \label{eq:eom_P}
 F_{\mu\nu}=H_{P} P_{\mu\nu} \qquad  \Rightarrow \qquad F=-\left(H_{P} \right)^{2} {P}.
 \end{align}
Inserting back this information in eq.~\eqref{eq:NLED_HP} and considering the inversion of the second equation which provides the dependence $P=P(F)$, we land in
  \begin{align} \label{eq:inv_legendre1}
 \mathcal{L}(F)= \left.\Big(2 H_{P} {P}-{H} \Big)\right|_{{P}={P}(F)}=
\frac{32 \alpha^3\lambda^{10}}{\kappa(\kappa \sqrt{2F} \pm \alpha\lambda)^2},
 \end{align}
where either of the plus or minus sign  choices is connected to the same  ${H}(P)$ electrodynamics. Notice that this form makes explicit the necessity of a nonlinear electrodynamics for the solution to exist. Namely, in a series expansion up to linear order, there appears a term proportional to $\sim \sqrt{2F}$, which cannot be independently turned off.


\end{document}